\documentclass[
reprint,
superscriptaddress,
amsmath,amssymb,
aps,
prb,
]{revtex4-2}
\usepackage[utf8]{inputenc}
\usepackage[english]{babel}
\usepackage[T1]{fontenc}
\usepackage[bookmarks=true,colorlinks,citecolor=red,linkcolor=red,allcolors=red]{hyperref}
\usepackage[capitalise]{cleveref}

\usepackage{csquotes}
\usepackage{physics}
\usepackage[toc,page]{appendix}

\usepackage{amsmath,amsfonts}
\usepackage{bm}

\usepackage[dvipsnames]{xcolor}

\usepackage{graphicx}

\begin{document}
\title{Non-stabilizerness of Neural Quantum States}

\author{Alessandro Sinibaldi}
\email{alessandro.sinibaldi@epfl.ch}
\affiliation{Institute of Physics, \'{E}cole Polytechnique F\'{e}d\'{e}rale de Lausanne (EPFL), CH-1015 Lausanne, Switzerland}
\affiliation{Center for Quantum Science and Engineering, EPFL, Lausanne, Switzerland}

\author{Antonio Francesco Mello}
\email{amello@sissa.it}
\affiliation{International School for Advanced Studies (SISSA), Via Bonomea 265, I-34136 Trieste, Italy}
\affiliation{Center for Computational Quantum Physics, Flatiron Institute, 162 5th Avenue, New York, NY 10010}

\author{Mario Collura}
\affiliation{International School for Advanced Studies (SISSA), Via Bonomea 265, I-34136 Trieste, Italy}
\affiliation{INFN, Sezione di Trieste, Via Valerio 2, 34127 Trieste, Italy}

\author{Giuseppe Carleo}
\affiliation{Institute of Physics, \'{E}cole Polytechnique F\'{e}d\'{e}rale de Lausanne (EPFL), CH-1015 Lausanne, Switzerland}
\affiliation{Center for Quantum Science and Engineering, EPFL, Lausanne, Switzerland}

\begin{abstract}
We introduce a methodology to estimate non-stabilizerness or ``magic'', a key resource for quantum complexity, with Neural Quantum States (NQS). 
Our framework relies on two schemes based on Monte Carlo sampling to quantify non-stabilizerness via Stabilizer Rényi Entropy (SRE) in arbitrary variational wave functions. 
When combined with NQS, this approach is effective for systems with strong correlations and in dimensions larger than one, unlike Tensor Network methods. 
Firstly, we study the magic content in an ensemble of random NQS, demonstrating that neural network parametrizations of the wave function capture finite non-stabilizerness besides large entanglement. 
Secondly, we investigate the non-stabilizerness in the ground state of the $J_1$-$J_2$ Heisenberg model. 
In 1D, we find that the SRE vanishes at the Majumdar-Ghosh point $J_2 = J_1/2$, consistent with a stabilizer ground state. 
In 2D, a dip in the SRE is observed near maximum frustration around $J_2/J_1 \approx 0.6$, suggesting a Valence Bond Solid between the two antiferromagnetic phases.
\end{abstract}

\maketitle

\paragraph*{Introduction. --}
Harnessing the exponential complexity of quantum systems to overcome the limitations of classical computing lies at the heart of quantum advantage. 
Entanglement has been identified as a fundamental feature that accounts for this complexity and has thus been thoroughly studied as a crucial resource~\cite{RevModPhys.82.277,cirac2012goals,houck2012chip}. 
However, it is also well known that entanglement is not the sole resource required to quantify quantum complexity. 
In particular, an equally essential resource is non-stabilizerness -- also known as ``magic''~\cite{gottesman1997stabilizer,gottesman1998heisenberg,aaronson2004improved}. 
The latter describes the degree to which a given quantum protocol diverges from polynomial simulability within the framework of the Gottesman-Knill theorem.

Although several measures of non-stabilizerness have been proposed~\cite{gross2007non,veitch2012negative,Gross2021}, evaluating most of these non-stabilizerness monotones remains computationally challenging in the realm of many-body systems~\cite{PhysRevLett.118.090501,bravyi2019simulation,heinrich2019robustness}. 
The underlying issue is the complexity of the minimization procedures required, which makes numerical evaluation intractable for systems larger than a few qubits. 
Recently, a significant advancement in overcoming this challenge has been marked by the development of more practical and computable measures. 
These include Bell magic~\cite{PRXQuantum.4.010301}, stabilizer nullity~\cite{Beverland2020}, and Stabilizer Rényi Entropies (SREs)~\cite{leone2022stabilizer}.
Remarkable progress has also been achieved in the experimental measurement of these quantities, marking a notable step forward in the practical assessment of non-stabilizerness~\cite{Oliviero2022,niroula2023phase,Bluvstein2023,bera2025nonstabilizernesssachdevyekitaevmodel,smith2024nonstabilizernesskineticallyconstrainedrydbergatom,dowling2025bridgingentanglementmagicresources}.
Among all these measures, SREs stand out for their favorable properties and computational tractability~\cite{leone2022stabilizer,haug2023stabilizer,leone2024stabilizer}, making them the central focus of this work. 

Despite these developments, much like the initial stages of understanding entanglement in many-body systems, the relationship between magic and physical phenomena (phase transition, criticality, etc.), as well as the interplay between magic and entanglement at the multi-particle level, remains largely unclear. 
A promising approach to address these questions lies in the use of Tensor Networks (TNs)~\cite{white1992density,orus2014practical,PhysRevLett.133.150604,mello2024clifforddressedtimedependentvariational}. 
They have currently offered the sole viable theoretical framework for computing magic on a large scale, employing both exact~\cite{haug2023quantifying,tarabunga2024nonstabilizerness,paviglianiti2024estimating} and stochastic methods~\cite{Lami2023,haug2023stabilizer,tarabunga2023manybody,tarabunga2023critical,PhysRevLett.133.010602}. 
TN methods are however limited to systems with weak entanglement and in low dimensionality.
In~\cite{spriggs1998quantum} the non-stabilizerness expressivity of some variational ansätze is examined for small system sizes.

In the present work we go beyond the TN realm, and extend the magic analysis to Neural Quantum States (NQS)~\cite{carleo2017solving}. Our contributions are threefold. 
First, we present methodologies to measure a SRE in quantum states encoded as NQS. 
Second, we apply these techniques to random NQS instances, showing that neural wave functions are able to capture a finite degree of non-stabilizerness besides extensive entanglement. 
Finally, we construct the magic phase diagram of the ground state of the 1D and 2D $J_1$-$J_2$ Heisenberg model, a paradigmatic quantum many-spin system that poses significant challenges for existing numerical methods. 
By leveraging the representative power of NQS ansätze, our approach effectively overcomes the limitations of state-of-the-art approaches for magic computation, enabling the investigation of non-stabilizerness in highly entangled many-body states in arbitrary spatial dimension. 
This opens new avenues for exploring the role of quantum resources in a plethora of physical scenarios.

\paragraph*{Non-stabilizerness resource theory. --}

Here we consider a quantum system composed of $N$ qubits (spins-$\frac{1}{2}$), extension to higher local dimension (qudits or larger spins) is straightforward. 
We define the set of Pauli matrices as $\{ \hat\sigma^{\mu} \}_{\mu=0}^3$, where $\hat\sigma^0 = \hat I$ is the identity matrix. 
The local computational basis $\{\ket{0}, \ket{1}\}$ is given by the eigenstates of the Pauli matrix $\hat{\sigma}^3$, such that $\hat{\sigma}^3 \ket{\sigma} = (-1)^\sigma \ket{\sigma}$ for $\sigma=0, 1$.
A general Pauli string $\hat{P}$ is identified by a set of indices $\boldsymbol{\mu}=\{\mu_1, \dots, \mu_N\}$, with $\mu_j\in\{0,1,2,3\}$ for $j=1, \ldots, N$, such that $\hat P(\boldsymbol{\mu}) = \bigotimes_{j=1}^N \hat\sigma^{\mu_j}_j \in \mathcal{P}_{N}$. Here  $\mathcal{P}_N$ encompasses all the possible Pauli strings with no overall phase.
The $N$-qubit Pauli group is therefore defined as 
$\mathcal{\tilde P}_{N} = \{\pm 1,\pm i\}\times\mathcal{P}_{N}$.

Let us now introduce another important group that allows us to identify the magic resource in quantum systems. 
Among all possible unitary transformations that can be applied to a system of $N$ qubits, there exists a special subset known as Clifford unitaries. 
These unitaries have the distinctive property of mapping Pauli strings to other Pauli strings and they form a group. 
The Clifford group is thus defined as the normalizer of the Pauli group.
It can be formally expressed as $\mathcal{C}_{N} = \{\hat{U} \,|\, \hat{U} \hat{P} \hat{U}^{\dagger} \in \mathcal{\tilde{P}}_{N}, \forall \, \hat{P} \in \mathcal{\tilde{P}}_{N} \}$ and can be generated using the Hadamard, phase, and CNOT gate.

With all the essential elements at hand, we can define the {\it stabilizer states}: these are states that can be built solely through Clifford operations applied to the state $|0\rangle^{\otimes N}$. 
These particular states, despite their intricate architecture and the substantial entanglement introduced by CNOT gates, do not provide any quantum advantage. 
The Gottesman-Knill theorem~\cite{gottesman1997stabilizer,gottesman1998heisenberg,aaronson2004improved} reveals all the limitations of stabilizer states, demonstrating that Clifford circuits can be efficiently simulated on classical computers. 
This revelation underscores the inherent constraints of stabilizer states in the realm of quantum computation.

In this framework, it is essential to quantify the non-Clifford resources required to prepare a specific quantum state. The degree of non-stabilizerness, often referred to as magic, needs to be evaluated using appropriate measures. 
For a pure possibly non-normalized state $|\Psi\rangle$, the Stabilizer R\'enyi Entropies (SREs)~\cite{leone2022stabilizer} are a possible choice. 
These are defined as:
\begin{equation}
\label{eq:SREs}
M_{\alpha} = \frac{1}{1-\alpha} \log \left[  \frac{1}{2^N}\sum_{\hat P \in \mathcal{P}_N} \bigg( \frac{\langle \Psi|\hat P|\Psi \rangle}{\langle \Psi|\Psi\rangle} \bigg)^{2\alpha} \right],
\end{equation}
where $\alpha$ is the R\'enyi index. 
If we now introduce the probability distribution over $\mathcal{P}_N$:
\begin{equation}\label{eq:classicalProb}
\Xi_{\hat P}=  \frac{1}{2^N}\left(\frac{\langle \Psi|\hat P|\Psi\rangle}{\langle \Psi|\Psi\rangle}\right)^2,
\end{equation}
the SREs can be written as averages over this probability distribution:
\begin{subequations}
\label{eq:pauli_sampling}
\begin{align}
  M_\alpha &= \frac{1}{1-\alpha} \log {\mathbb{E}}_{\hat P \sim \Xi_{\hat P}}  [\Xi_{\hat P}^{\alpha-1}]  -N\log 2, &\text{for}\,\,\, \alpha\neq 1, \\
  M_1 &=   -{\mathbb{E}}_{\hat P \sim \Xi_{\hat P}}[\log \Xi_{\hat P} ] - N\log 2.
\end{align}
\end{subequations}

Recent research has confirmed that for qubits and pure states, the SREs with $\alpha \geq 2$ exhibit monotonic behavior~\cite{leone2024stabilizer}, meaning that they serve as valid measures of non-stabilizerness in a resource theory perspective. On the contrary, it has been observed that the monotonicity does not hold for $0 \leq \alpha < 2$, where violations are evident~\cite{haug2023stabilizer}.
Therefore, in the following we consider $\alpha = 2$. 

Multiple ways to efficiently estimate the SREs have been proposed, but each of these has a restricted validity.  
When the many-body quantum state $\ket{\Psi}$ can be effectively encoded as a Matrix Product State (MPS), there exist algorithms to calculate $M_\alpha$ directly from~\cref{eq:SREs}~\cite{haug2023quantifying,haug2023stabilizer,tarabunga2024nonstabilizerness}.
These schemes inherit all the limitations of TN approaches, namely they are efficiently applicable to low-entanglement systems and one-dimensional settings only.
To overcome these limits, an approach based on Quantum Monte Carlo (QMC) simulation has been proposed~\cite{liu2024non}.
While this QMC method can handle systems in dimensions larger than one, its applicability is still constrained to wave functions that do not exhibit a sign problem~\cite{troyer2005computational}.
Recently, an effective method for assessing non-stabilizerness in fermionic Gaussian states has also been introduced~\cite{collura2025quantummagicfermionicgaussian}.
The most general approach to compute SREs remains however estimating them from~\cref{eq:pauli_sampling}, by performing a sampling in the space of Pauli strings $\mathcal{P}_N$ according to the probability distribution $\Xi_{\hat P}$ in~\cref{eq:classicalProb}.
In the special case of an MPS ansatz, perfect sampling from exactly computable conditional probabilities can be performed~\cite{Lami2023,PhysRevLett.133.010602}.
For a generic quantum state with query and sampling access~\cite{best2010simulating}, one has to resort to Markov chain Monte Carlo in $\mathcal{P}_N$, as proposed in~\cite{tarabunga2023manybody}. 

In this work, we present two methods based on Monte Carlo sampling of the wave function to estimate non-stabilizerness in arbitrary many-qubit systems.
We call these two approaches \emph{replicated estimator} and \emph{Bell basis estimator}.
In all the following, we denote as $\mathcal{H}$ the $N$-qubit Hilbert space and as $\ket{\Psi}$ the 
state for which we measure the non-stabilizerness. 

\begin{figure}
    \centering
    \includegraphics[width=1.0\linewidth]{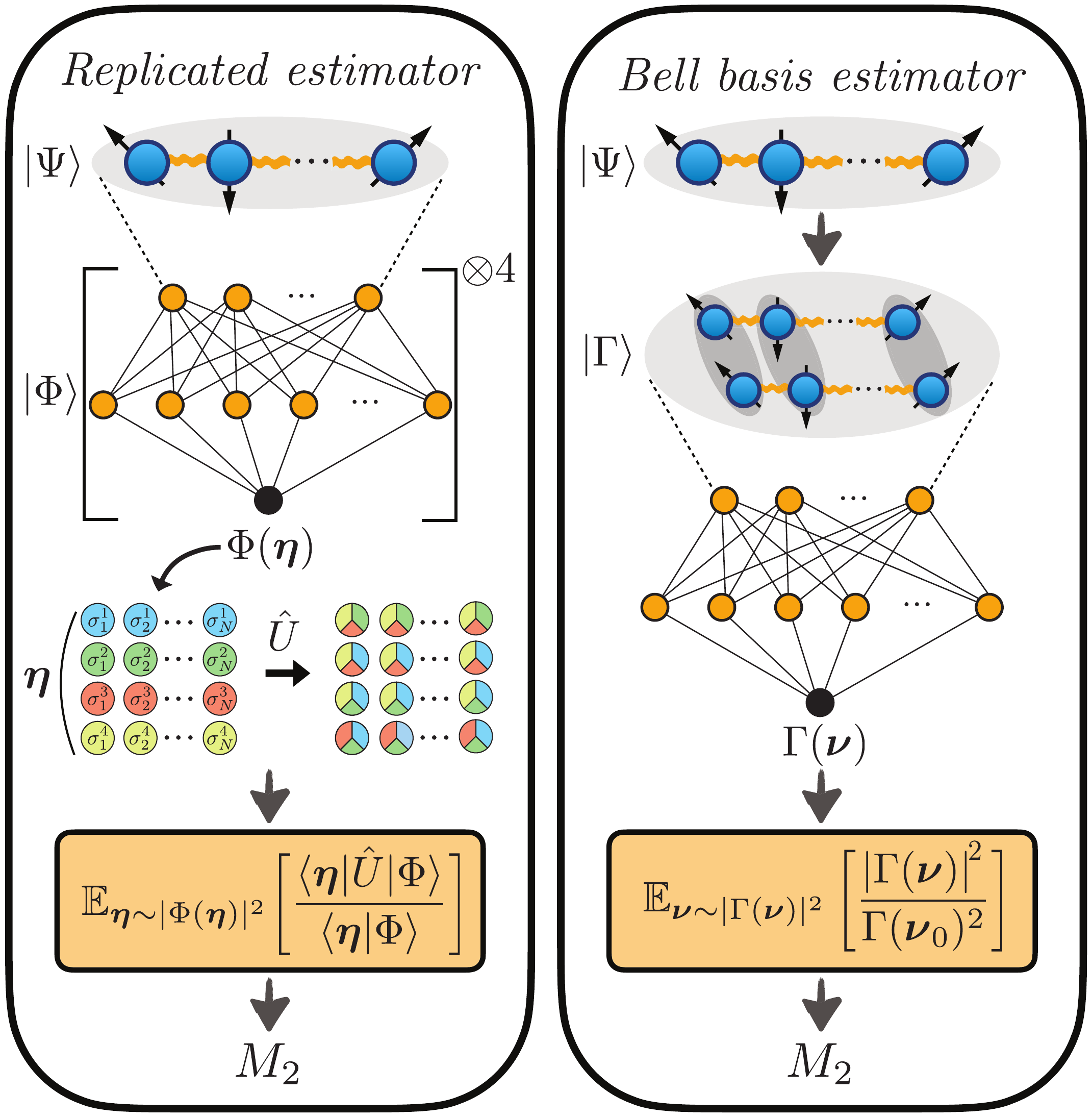}
    \caption{
    Sketch of the two approaches for estimating the SRE $M_2$ of a many-qubit quantum state $\ket{\Psi}$ through Monte Carlo sampling.
    The replicated estimator consists of considering four unentangled copies of $\ket{\Psi}$ and sampling the expectation value of an operator $\hat{U}$ which creates interaction between the replicas. 
    The Bell basis estimator involves two entangled copies of $\ket{\Psi}$ and computes the non-stabilizerness by averaging amplitude ratios.
    Each copy $\ket{\Psi}$ in the quadrupled state $\ket{\Phi}$ and the entire doubled entangled state $\ket{\Gamma}$ are encoded as NQS. 
    The stochastic estimates of the $M_2$ are performed by Monte Carlo sampling from the Born distributions of $\ket{\Phi}$ and $\ket{\Gamma}$. 
    }
    \label{fig:sketch}
\end{figure}

\paragraph*{Replicated estimator. --}
Consider the four times replicated Hilbert space $\bigotimes_{i=1}^4 \mathcal{H}$ with configurations $|\boldsymbol{\eta}\rangle =    |\sigma^{(1)},\sigma^{(2)},\sigma^{(3)},\sigma^{(4)}\rangle\equiv|\sigma^{(1)}\rangle\otimes|\sigma^{(2)}\rangle\otimes|\sigma^{(3)}\rangle\otimes|\sigma^{(4)}\rangle$, where $\ket*{\sigma^{(j)}} \in \mathcal{H}$ for $j=1, \ldots, 4$. 
Only for this estimator, we adopt the local spin basis $\{-1, 1\}$ instead of $\{0, 1\}$.
Moreover, consider the quantum state:
\begin{equation}
|\Phi\rangle	=	|\Psi^{*},\Psi^{*},\Psi^{*},\Psi\rangle,
\end{equation}
and the unitary operator defined by: 
\begin{equation}
\begin{split}
\label{eq:transformation_U}
\hat{U}|\boldsymbol{\eta}\rangle = |&\sigma^{(2)}\odot\sigma^{(3)}\odot\sigma^{(4)}, \sigma^{(1)}\odot\sigma^{(3)}\odot\sigma^{(4)}, \\
&\sigma^{(1)}\odot\sigma^{(2)}\odot\sigma^{(4)},\sigma^{(1)}\odot\sigma^{(2)}\odot\sigma^{(3)}\rangle,
\end{split}
\end{equation}
where $\odot$ denotes the Hadamard product between the qubit strings. 
Using the results of~\cite{tarabunga2023magic}, one can show that:
\begin{equation}
\label{eq:replicated_estimator}
\exp(-M_{2}) = \frac{\langle\Phi|\hat{U}|\Phi\rangle}{\langle\Phi|\Phi\rangle} = \mathbb{E}_{\boldsymbol{\eta} \sim |\Phi(\boldsymbol{\eta})|^2}\bigg[\frac{\bra{\boldsymbol{\eta}} \hat{U} \ket{\Phi}}{\bra{\boldsymbol{\eta}}\ket{\Phi}}\bigg].
\end{equation}

Therefore, $M_2$ can be computed from an expectation value over $|\Phi\rangle$, which can be efficiently estimated by sampling from the distribution $|\langle \boldsymbol{\eta}|\Phi\rangle|^2/\langle \Phi | \Phi \rangle$. 
The details are reported in the Supplemental Material~\cite{suppmat}.

\paragraph*{Bell basis estimator. --}
The second method for computing non-stabilizerness is based on the equivalence between the expectation value of a Pauli string and the amplitude over a Bell state of two copies of the original system~\cite{montanaro2017learning,Gross2021,9714418,PRXQuantum.4.010301,PhysRevLett.132.240602}.
This estimator can be used only when $\ket{\Psi}$ is the ground state of an Hamiltonian $\hat H$.
As shown in the Supplemental Material~\cite{suppmat}, we can write:
\begin{equation}
\label{eq:bell_basis_estimator}
    \exp(-M_2) =
     \mathbb{E}_{\boldsymbol{\nu} \sim |\Gamma(\boldsymbol{\nu})|^2}\left[\frac{|\Gamma(\boldsymbol{\nu})|^2}{\Gamma(\boldsymbol{\nu}_0)^2}\right],
\end{equation}
where $\ket{\boldsymbol{\nu}}, \ket{\boldsymbol{\nu}_0} \in \mathcal{H} \otimes \mathcal{H}$ with $\ket{\boldsymbol{\nu}_0} = \ket{0, 0, \ldots, 0} $ and $|\Gamma\rangle$ is the ground state of the transformed Hamiltonian $\hat{\mathbb{H}} = \hat{C}^{\dag} (\hat{H} \otimes \hat{I} + \hat{I} \otimes \hat{H}^*) \hat{C}$, with $\hat{I}$ the identity and $\hat C$ a suitable Clifford transformation entangling the two copies of the system.
In the same way as for~\cref{eq:replicated_estimator}, the average in~\cref{eq:bell_basis_estimator} can be estimated via Monte Carlo sampling from $|\langle \boldsymbol{\nu}|\Gamma\rangle|^2/\langle \Gamma | \Gamma \rangle$. 

In our framework, the states $\ket{\Phi}$ and $\ket{\Gamma}$ are encoded as suitable NQS. 
In general, the replicated estimator suffers from larger statistical fluctuations than the Bell basis estimator. 
This occurs because $\hat{U}$ in~\cref{eq:transformation_U} is a non-local transformation, so $\bra{\boldsymbol{\eta}} \hat{U} \ket{\Phi}$ and $\bra{\boldsymbol{\eta}}\ket{\Phi}$ can differ significantly, resulting in a stochastic estimate that is affected by statistical outliers.
This is the same issue occurring in the Monte Carlo computation of the Rényi-2 entanglement entropy~\cite{HastingsPRL2010}.
This problem, which is more severe for large system sizes, can be mitigated by increasing the number of effective samples with U-statistics, as done in~\cite{falcao2024non}, or by employing a scheme inspired to annealed importance sampling~\cite{neal2001annealed}, detailed in the Supplemental Material~\cite{suppmat}. 
The Bell basis estimator is free from such a problem, since for it the distribution used in the sampling is proportional to the estimator itself.
The sample complexities of the two estimators can be analyzed from the scaling of the statistical errors on $M_2$ with the system size $N$. 
As detailed in the Supplemental Material~\cite{suppmat}, the statistical errors are: 
\begin{align}
    \label{eq:err_repl}
    \Delta M_2^{\text{repl.}} &= \sqrt{\frac{e^{2 M_2} - 1}{N_s}}, \\
    \label{eq:err_bell}
    \Delta M_2^{\text{Bell}} &= \sqrt{\frac{e^{2 (M_2 - M_3)} - 1}{N_s}}, 
\end{align}
where $N_s$ is the number of samples used in the empirical averages. 
We note that, since the SREs satisfy the inequalities $M_\alpha \geq M_\beta$ for $\alpha < \beta$, the exponent in~\cref{eq:err_bell} is positive.
\cref{eq:err_repl,eq:err_bell} entail that for both the estimators, if the SREs grow at most logarithmically with $N$, the number of samples $N_s$ required to keep the errors fixed scales polynomial with $N$, meaning that the estimators are efficient. 
Logarithmic growth of SREs can arise in certain frustrated many-body systems~\cite{odavic2023complexity}. 
Conversely, when the SREs scale linearly with $N$ -- as is typical for generic quantum many-body states -- the numerators in~\cref{eq:err_repl,eq:err_bell} grow exponentially with $N$, and so does the required number of samples $N_s$.
Nevertheless, the sample complexity still grows substantially slower than the cost of an exact computation, which scales as $2^{2N}$.
In practice, these estimators remain valuable as they enable the exploration of system sizes that are otherwise intractable with exact methods.
This is the same complexity as that of the Pauli-Markov chain method for the SREs~\cite{tarabunga2023manybody} and the Monte Carlo estimation of the Rényi-2 entanglement entropy~\cite{HastingsPRL2010} for volume-law quantum states.
    
We remark that, while the replicated estimator is in general non-positive, the Bell basis estimator is instead positive-definite by construction and thus more robust for wave functions with sign structure. 
The Bell basis estimator, however, is limited to ground states and requires knowing the typically highly-entangled state $\ket{\Gamma}$, which becomes challenging for complicated models.  
Even when the physical Hamiltonian $\hat{H}$ contains only local interactions, the Clifford transformation $\hat{C}$ introduces non-local couplings between the qubits of the system and those of the replica.
Additionally, the interactions typically become more complex, with originally simple two-body terms in $\hat{H}$ transforming into higher-order interactions in $\hat{\mathbb{H}}$, often involving four-body couplings.
This severely complicates learning the ground state of $\hat{\mathbb{H}}$ through Variational Monte Carlo.
Since these limitations are constraining for the applications we chose, we adopt the replicated estimator in all our calculations shown in the main text. 
The sketch in Fig.~\ref{fig:sketch} pictorially illustrates the two approaches. 
In the Supplemental Material~\cite{suppmat} we benchmark the two Monte Carlo estimators against a reference simulation performed with MPS.

\paragraph*{Ensemble of random NQS. --}
As a first application of the methods, we investigate the representative power of NQS in terms of non-stabilizerness. 
To do so, we analyze the magic content of an ensemble of NQS with random parameters. 
We consider the ensemble studied in~\cite{PhysRevX.7.021021} for entanglement properties, namely a set of Restricted Boltzmann Machines (RBM)~\cite{carleo2017solving} for $N$ spins-$\frac{1}{2}$: 
\begin{equation}
    \Psi_{\theta}(\sigma) = e^{\sum_{i=1}^{N} a_i \sigma_i} \prod_{i=1}^{M} \cosh(\sum_{j=1}^{N} W_{ij} \sigma_j + b_i), 
\end{equation}
where the parameters $\theta=(a, b, W)$ are chosen such that the visible and hidden biases $a$ and $b$ are set to zero, while the complex weights $W_{ij}$ are drawn uniformly from $[-3/N, 3/N]$ in the real part and from $[-\pi, \pi]$ in the imaginary part.
The density of hidden neurons $\alpha = M/N$ is fixed to $1$, since for this choice the average von Neumann entanglement entropy of the ensemble obeys a volume-law, as shown in Ref.~\cite{PhysRevX.7.021021}.
We compute the $M_2$ averaged over several independent realizations of these random RBM for different numbers of spins $N$. 

\begin{figure}[h]
    \centering
    \includegraphics[width=1.0\linewidth]{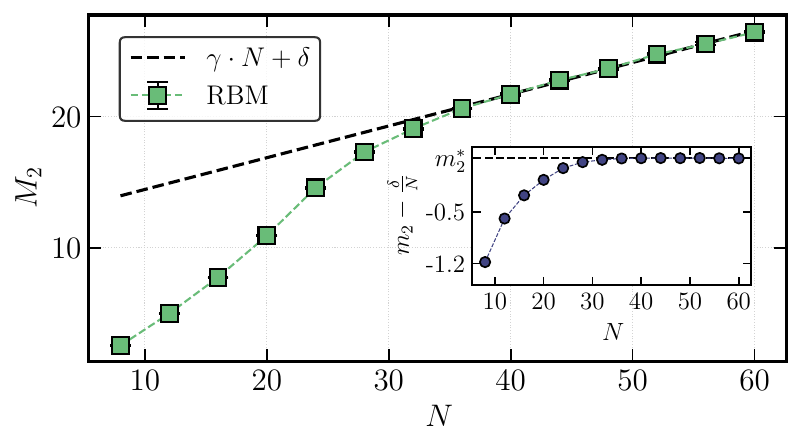}
    \caption{Averaged SRE $M_2$ in the ensemble of random RBM for various system sizes $N$. 
    Each point is the mean over $10^3$ independent realizations of the ensemble. 
    The values of the $M_2$ are estimated using $N_s = 2^{22} \approx 4 \cdot 10^6$ samples. 
    The dashed line corresponds to a linear fit on the last $7$ data points. 
    The inset displays the SRE density $m_2=M_2/N$ without the offset coming from the intercept of the fit.
    The dashed line in the inset indicates the asymptotic value $m^*_2$ of the SRE density for $N \rightarrow \infty$ which coincides with the slope of the fit.
    }
    \label{fig:random_rbms}
\end{figure}

\cref{fig:random_rbms} shows that for  an enough large system size the averaged $M_2$ of the ensemble increases linearly with $N$, as confirmed by the linear fit. 
Consequently, as displayed in the inset, the SRE density $m_2=M_2/N$ converges to a finite value in the thermodynamic limit, which corresponds to the slope of the fit since $m_2 = \gamma + \delta/N$. 
The asymptotic value is estimated as $m_2^* = \gamma = 0.241 \pm 0.005$ and quantifies the magic content in this ensemble of random NQS.
This calculation shows that NQS parametrizations are capable of capturing finite non-stabilizerness besides large entanglement, simultaneously accounting for two distinct types of quantum resources.

\begin{figure}[h]
    \centering
    \includegraphics[width=1.0\linewidth]{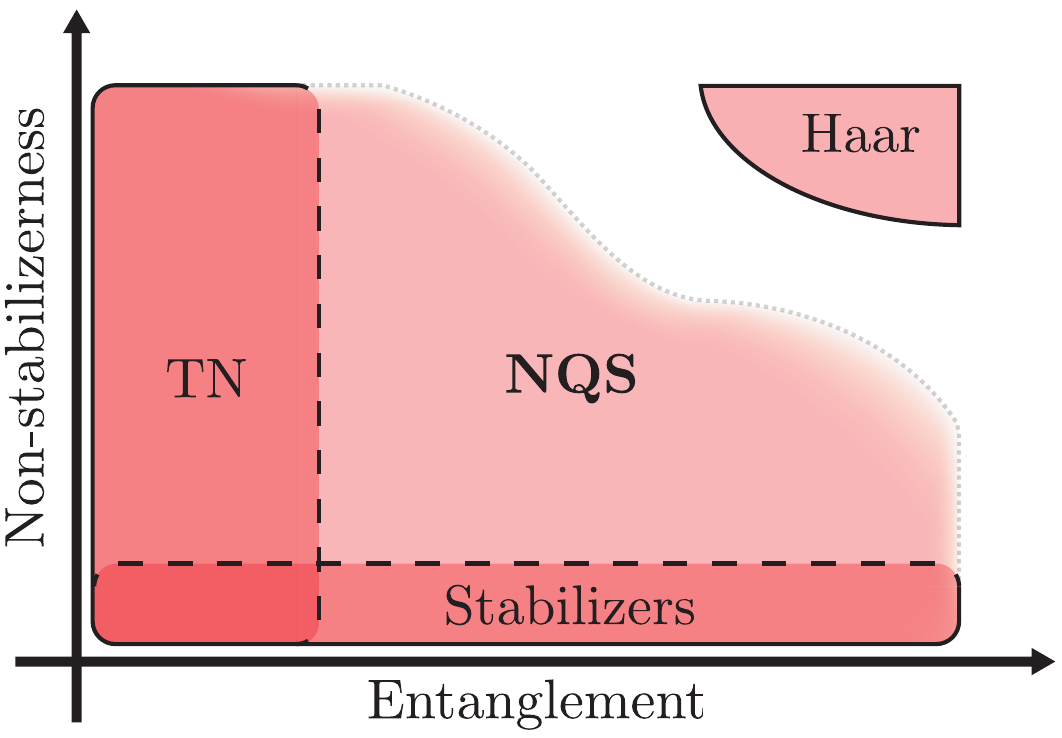}
    \caption{
    Pictorial diagram illustrating the expressive power of some Amplitude Ratio quantum states~\cite{havlicek2023amplitude} in terms of entanglement and non-stabilizerness.
    Here TN indicates all the planar Tensor Network ansätze, including MPS as well as Projected Entangled Pair States~\cite{verstraete2004renormalization}.
    }
    \label{fig:magic_vs_entanglement}
\end{figure}

The scheme in~\cref{fig:magic_vs_entanglement} illustrates the representative power of NQS in terms of entanglement and magic content compared to other states of the Amplitude Ratio (AR) family~\cite{havlicek2023amplitude}. 
This family comprises all states, generally non-normalized, from which one can sample and compute exponentially accurate amplitude ratios with polynomial complexity.
TN encompass states with low entanglement but arbitrarily high non-stabilizerness~\cite{lami2024quantum}, while stabilizer states have no magic but are possibly highly-entangled. 
NQS are known to represent any efficiently-contracted TN wave function~\cite{sharir2022neural,wu2023tensor} and any stabilizer state~\cite{zhang2018efficient,jia2019efficient,lu2019efficient,zheng2019restricted,pei2021compact}.
Additionally, NQS can encode states with volume-law entanglement~\cite{PhysRevX.7.021021,denis2023comment} and finite non-stabilizerness, as proved previously, as well as states with moderate entanglement but high magic content, as shown in the next section. 
Therefore, NQS allow us to represent AR states very close to random Haar states, which have nearly maximal values of both entanglement~\cite{PhysRevLett.71.1291} and non-stabilizerness~\cite{lami2024quantum}.
We note that the upper boundaries of NQS in~\cref{fig:magic_vs_entanglement} are speculative, as suggested by the faded edge, since it is still unknown if NQS can represent random Haar states.

\paragraph*{$J_1$-$J_2$ Heisenberg model. --}
We now explore the non-stabilizerness in the ground state of a paradigmatic quantum many-body spin system, the $J_1$-$J_2$ Heisenberg model with Hamiltonian: 
\begin{equation}
\label{eq:j1j2}
    \hat{H} = J_1 \sum_{\langle i, j \rangle} \hat{\boldsymbol{S}}_{i} \cdot \hat{\boldsymbol{S}}_j + J_2 \sum_{\langle \langle i, j \rangle \rangle} \hat{\boldsymbol{S}}_{i} \cdot \hat{\boldsymbol{S}}_j, 
\end{equation}
where $\hat{\boldsymbol{S}}_i = (\hat{S}_i^x, \hat{S}_i^y, \hat{S}_i^z)$ is the spin-$\frac{1}{2}$ operator at site $i$, and $J_1, J_2 > 0$ are the antiferromagnetic couplings between nearest- and next-nearest-neighbor spins respectively. 
We consider both a 1D spin chain and a 2D square lattice with periodic boundary conditions.
The physical properties of the ground state of~\cref{eq:j1j2} have been extensively investigated over the years, with particular focus on the regime of high frustration which revealed to be challenging for numerical methods~\cite{eggert1996numerical,white1996dimerization,hu2013direct,PhysRevLett.113.027201,morita2015quantum,liang2018solving,wang2018critical,haghshenas2018u,ferrari2019neural,choo2019two,szabo2020neural,liang2021hybrid,nomura2021dirac,viteritti2022accuracy,chen2022systematic,li2022bridging,roth2023high,liang2023deep,ledinauskas2023scalable,reh2023optimizing,viteritti2023transformer_1d,chen2024empowering,rende2024simple,wang2024variational}.
It is therefore of particular interest to analyze the ground state in terms of non-stabilizer resources.
Owing to the large entanglement of some phases and the possible high-dimensionality, standard TN approaches to compute non-stabilizerness~\cite{Lami2023,haug2023stabilizer,tarabunga2023manybody,tarabunga2023critical,haug2023quantifying,tarabunga2024nonstabilizerness,paviglianiti2024estimating,PhysRevLett.133.010602} are impractical.
On the other side, QMC-based schemes~\cite{liu2024non} are not applicable due to the non-trivial sign structure of the ground state wave function for arbitrary couplings. 

\begin{figure}
    \centering
    \includegraphics[width=1.0\linewidth]{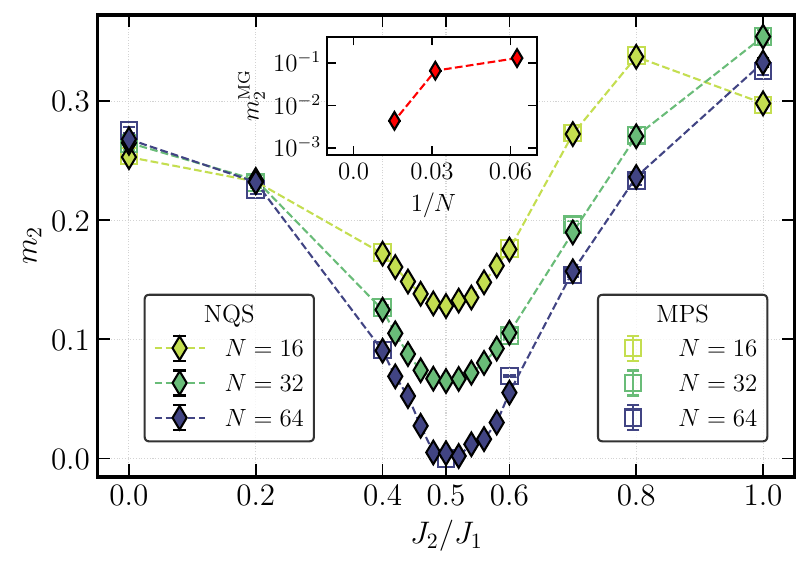}
    \caption{SRE density $m_2$ in the ground state of the 1D $J_1$-$J_2$ Heisenberg model for different values of $J_2/J_1$. 
    Results are shown for chains with $N=16, 32, 64$ spins.
    The values of the SRE density are estimated using $N_s \approx 10^{9}$ samples. 
    The inset shows the size scaling of the $m_2$ at the Majumdar-Ghosh point $J_2 = J_1/2$, denoted as $m_2^{\text{MG}}$.
    The ground state is approximated using a complex RBM with $\alpha=4$ for $N=16$ and the ViT for $N=32, 64$.
    The results from perfect sampling with MPS~\cite{Lami2023} using bond dimension $\chi=400$ and $N_s =10^4$ samples are reported as a benchmark. 
    }
    \label{fig:j1j2_1d}
\end{figure}

\begin{figure}
    \centering
    \includegraphics[width=1.0\linewidth]{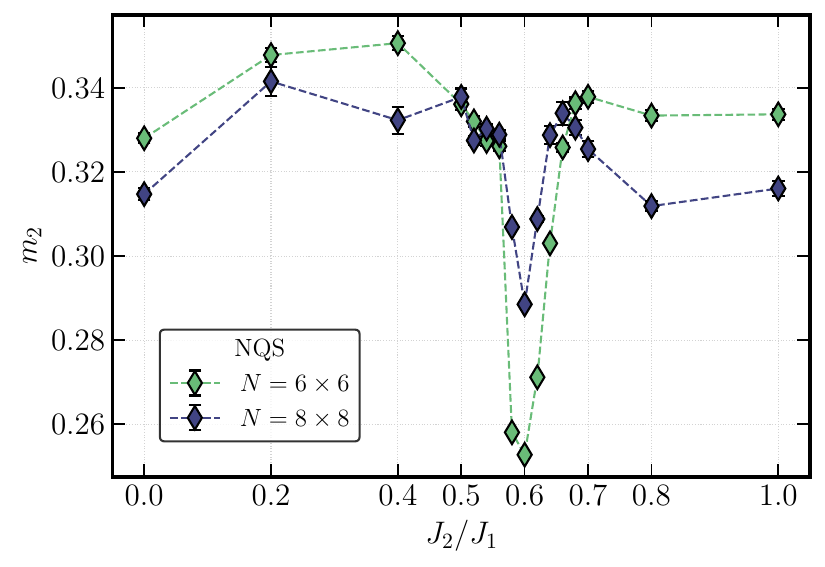}
    \caption{SRE density $m_2$ in the ground state of the 2D $J_1$-$J_2$ Heisenberg model for different values of $J_2/J_1$. 
    Results are shown for the $6 \times 6$ and the $8 \times 8$ lattices.
    The values of the SRE density are estimated using $N_s \approx 4 \cdot 10^{6}$ samples. 
    The ground state is approximated by employing the ViT architecture.}
    \label{fig:j1j2_2d}
\end{figure}

We approximate the ground state of~\cref{eq:j1j2} for different values of $J_2/J_1$ using Variational Monte Carlo (VMC) with a powerful NQS parametrization of the wave function. 
Specifically, we employ the Deep Vision Transformer (ViT) architecture introduced in~\cite{viteritti2023transformer_1d, viteritti2023transformer_2d} together with the modern formulation of the Stochastic Reconfiguration ~\cite{sorella1998green,sorella2005wave} method in the limit of numerous variational parameters~\cite{rende2024simple,chen2024empowering}. 
Some details on the ansatz and the optimization protocol are reported in the Supplemental Material~\cite{suppmat}.
Once we obtain the ground states, we compute the $M_2$ for each of them. 

Since we cannot access the exact wave function for the considered system sizes, it is imperative to quantify the variational error committed on the SRE due to the NQS approximation of the ground state.
The calculation for $m_2$ is shown in the Supplemental Material~\cite{suppmat}, and it follows that its variational error is estimated at leading order as $\sqrt{\text{Var}(\hat{H})} / N$, where $\text{Var}(\hat{H})$ is the energy variance on the optimized NQS.
From the Monte Carlo calculation of the variance, we can therefore estimate the systematic error on $m_2$ due to the variational approximation. 

\cref{fig:j1j2_1d} reveals that in the 1D system the $m_2$ attains a minimum in correspondence of $J_2=J_1/2$, namely when the 1D $J_1$-$J_2$ Heisenberg model corresponds to the exactly solvable Majumdar-Ghosh (MG) model~\cite{majumdar1969next}. 
In particular, the minimum of the SRE density approaches zero as the system size increases, as shown in the inset.
This is in perfect agreement with what is expected from the MG ground state, since it is known that this consists of a valence bond state composed by singlets of maximally entangled spins, which is a stabilizer. 
With periodic boundary conditions, the actual ground state is in general a superposition of two degenerate ground states which differ by translating the singlets by one lattice spacing. 
Therefore, the non-stabilizerness can be small yet not-vanishing at finite size.
However, as we approach the thermodynamic limit, the degeneracy disappears and the ground state becomes a single stabilizer. 
The results from MPS calculations~\cite{Lami2023} closely match those from NQS across the full phase diagram, confirming the accuracy of our findings for the 1D system. 

For the 2D model, as displayed in~\cref{fig:j1j2_2d}, the $m_2$ shows a clear dip in the regime of maximum frustration reaching a finite minimum around $J_2/J_1 \approx 0.6$.
This behavior is reminiscent of the 1D case in~\cref{fig:j1j2_1d}, but differs since the minimal $m_2$ remains non-vanishing as we increase the system size. 
What is observed in the 2D model around $J_2/J_1 \approx 0.6$ is compatible with a Valence Bond Solid (VBS) phase up to some finite-size correction, which implies a reduction of the non-stabilizerness without complete cancellation. 
The presence of the VBS between the Néel (small $J_2/J_1$) and the stripe-type (large $J_2/J_1$) antiferromagnetic orders in the 2D $J_1$-$J_2$ model have been proposed in several previous works~\cite{PhysRevLett.113.027201,morita2015quantum,wang2018critical,haghshenas2018u,nomura2021dirac}.
For the 2D system, TN cannot serve as a reliable benchmark, as NQS have demonstrated superior accuracy in capturing the ground state~\cite{rende2024simple}.
Even when accurate TN approximations are attainable~\cite{PhysRevB.109.L161103,PhysRevLett.113.027201}, the large bond dimension $\chi$ required renders the computation of the SRE practically infeasible. 
Indeed, the scaling with $\chi$ of TN-based algorithms for estimating the $M_2$, such as the MPS perfect sampling~\cite{Lami2023} with $O(N_s N \chi^3)$, the replica approach~\cite{haug2023quantifying} with $O(N \chi^{12})$ and the Pauli-Markov chain method~\cite{tarabunga2023manybody} with $O(\log N \chi^4)$, limits computations to small bond dimensions, for which the ground state approximation becomes inadequate. 

\paragraph*{Outlook and Conclusions. --}
In this work, we introduce a framework for quantifying non-stabilizerness in terms of Stabilizer Rényi Entropies (SREs) with Neural Quantum States (NQS). 
We present two Monte Carlo algorithms to measure non-stabilizerness in quantum many-body states encoded in arbitrary variational ansätze.
Combining these schemes with NQS enables overcoming the limitations of TN methods imposed by extensive correlations and high dimensionality.
Firstly, we characterize the magic content of random NQS wave functions, proving that neural network parametrizations are able to capture multiple quantum resources simultaneously, namely both finite non-stabilizerness and large entanglement. 
Secondly, we compute the SRE in the ground state of the $J_1$-$J_2$ Heisenberg model on the spin chain and on the square lattice. 
This frustrated quantum spin system cannot be efficiently addressed with state-of-the-art approaches for magic computation. 
Our simulations show that in 1D the non-stabilizerness decreases to zero at $J_2=J_1/2$  in the thermodynamic limit, which is in agreement with the stabilizer nature of the Majumdar-Ghosh ground state. 
In 2D, instead, the SRE displays a clear dip in the regime of maximum frustration with a non-vanishing minimum around $J_2/J_1 \approx 0.6$.
This diminution in the non-stabilizerness aligns with the presence of a Valence Bond Solid phase in the intermediate region between the two antiferromagnetic ordered regimes.

A potential development of this work is extending the magic analysis to interacting fermionic systems. 
Additionally, an important direction is the investigation of the non-stabilizerness dynamics in the time evolution of strongly-correlated quantum systems simulated with NQS methods~\cite{carleo2017solving,schmitt2020quantum,sinibaldi2023unbiasing,gravina2024neural,sinibaldi2024time,nys2024ab}.
Finally, it is relevant to thoroughly assess the capability of NQS to represent states with maximal quantum complexity, namely those that simultaneously saturate both entanglement and non-stabilizerness, akin to random Haar states. 

\section{Software}
The code for the simulations is based on the software package NetKet~\cite{netket2,vicentini2022netket} and is available at~\cite{repo}. 
We acknowledge the use of Stim~\cite{gidney2021stim} for the stabilizer formalism operations, and of ITensor~\cite{itensor22} for the MPS calculations.

\section{Acknowledgments}
We warmly thank V. Savona, M. Dalmonte, P. Tarabunga and G. Lami for inspirational and insightful discussions. 
We gratefully acknowledge Luciano Loris Viteritti and Riccardo Rende for providing the code of the ViT ansatz and assisting in its VMC optimization.  
A. S. is supported by SEFRI under Grant No.\ MB22.00051 (NEQS - Neural Quantum Simulation). 
The Flatiron Institute is a division of the Simons Foundation.
M. C. thanks support by the PNRR MUR project PE0000023-NQSTI, and the PRIN 2022 (2022R35ZBF) - PE2 - ``ManyQLowD''.

\nocite{mbeng2024quantum,oliviero2022ising,liu2022convnet,bengio2014representationlearningreviewnew,denis2024accurate,marshall1955antiferromagnetism}

\bibliography{bibliography}

@article{cirac2012goals,
    doi = {10.1038/nphys2275},
  url = {https://doi.org/10.1038/nphys2275},
  year = {2012},
  month = apr,
  publisher = {Springer Science and Business Media {LLC}},
  volume = {8},
  number = {4},
  pages = {264--266},
  author = {J. Ignacio Cirac and Peter Zoller},
  title = {Goals and opportunities in quantum simulation},
  journal = {Nature Physics}
}

@article{houck2012chip,
   doi = {10.1038/nphys2251},
  url = {https://doi.org/10.1038/nphys2251},
  year = {2012},
  month = apr,
  publisher = {Springer Science and Business Media {LLC}},
  volume = {8},
  number = {4},
  pages = {292--299},
  author = {Andrew A. Houck and Hakan E. T\"{u}reci and Jens Koch},
  title = {On-chip quantum simulation with superconducting circuits},
  journal = {Nature Physics}
}

@misc{gottesman1997stabilizer,
      title={{Stabilizer Codes and Quantum Error Correction}}, 
      author={Daniel Gottesman},
      year={1997},
      eprint={quant-ph/9705052},
      archivePrefix={arXiv},
}

@article{aaronson2004improved,
  title = {Improved simulation of stabilizer circuits},
  author = {Aaronson, Scott and Gottesman, Daniel},
  journal = {Phys. Rev. A},
  volume = {70},
  issue = {5},
  pages = {052328},
  numpages = {14},
  year = {2004},
  month = {Nov},
  publisher = {American Physical Society},
  doi = {10.1103/PhysRevA.70.052328},
  url = {https://link.aps.org/doi/10.1103/PhysRevA.70.052328}
}

@misc{gottesman1998heisenberg,
      title={{The Heisenberg Representation of Quantum Computers}},
      author={Gottesman, Daniel},
      year={1998},
      eprint={quant-ph/9807006},
      archivePrefix={arXiv},
}

@article{bravyi2019simulation,
  title = {Simulation of quantum circuits by low-rank stabilizer decompositions},
  volume = {3},
  ISSN = {2521-327X},
  url = {http://dx.doi.org/10.22331/q-2019-09-02-181},
  DOI = {10.22331/q-2019-09-02-181},
  journal = {Quantum},
  publisher = {Verein zur Forderung des Open Access Publizierens in den Quantenwissenschaften},
  author = {Bravyi,  Sergey and Browne,  Dan and Calpin,  Padraic and Campbell,  Earl and Gosset,  David and Howard,  Mark},
  year = {2019},
  month = sep,
  pages = {181}
}

@article{heinrich2019robustness,
   doi = {10.22331/q-2019-04-08-132},
  url = {https://doi.org/10.22331/q-2019-04-08-132},
  year = {2019},
  month = apr,
  publisher = {Verein zur Forderung des Open Access Publizierens in den Quantenwissenschaften},
  volume = {3},
  pages = {132},
  author = {Markus Heinrich and David Gross},
  title = {{Robustness of Magic and Symmetries of the Stabiliser Polytope}},
  journal = {Quantum}
}

@article{veitch2012negative,
  title={Negative quasi-probability as a resource for quantum computation},
  author={Veitch, Victor and Ferrie, Christopher and Gross, David and Emerson, Joseph},
  journal={New Journal of Physics},
  volume={14},
  number={11},
  pages={113011},
  year={2012},
  publisher={IOP Publishing},
  doi={10.1088/1367-2630/14/11/113011},
  url={https://iopscience.iop.org/article/10.1088/1367-2630/14/11/113011}
}

@article{gross2007non,
  doi = {10.1007/s00340-006-2510-9},
  url = {https://doi.org/10.1007/s00340-006-2510-9},
  year = {2006},
  month = dec,
  publisher = {Springer Science and Business Media {LLC}},
  volume = {86},
  number = {3},
  pages = {367--370},
  author = {D. Gross},
  title = {{Non-negative Wigner functions in prime dimensions}},
  journal = {Applied Physics B}
}

@article{itensor22,
	title={{The ITensor Software Library for Tensor Network Calculations}},
	author={Matthew Fishman and Steven R. White and E. Miles Stoudenmire},
	journal={SciPost Phys. Codebases},
	pages={4},
	year={2022},
	publisher={SciPost},
	doi={10.21468/SciPostPhysCodeb.4},
	url={https://scipost.org/10.21468/SciPostPhysCodeb.4}
}

@article{leone2022stabilizer,
    title = {{Stabilizer Rényi Entropy}},
  author = {Leone, Lorenzo and Oliviero, Salvatore F. E. and Hamma, Alioscia},
  journal = {Phys. Rev. Lett.},
  volume = {128},
  issue = {5},
  pages = {050402},
  numpages = {5},
  year = {2022},
  month = {Feb},
  publisher = {American Physical Society},
  doi = {10.1103/PhysRevLett.128.050402},
  url = {https://link.aps.org/doi/10.1103/PhysRevLett.128.050402}
}

@article{PhysRevLett.132.240602,
  title = {{Efficient Quantum Algorithms for Stabilizer Entropies}},
  author = {Haug, Tobias and Lee, Soovin and Kim, M. S.},
  journal = {Phys. Rev. Lett.},
  volume = {132},
  issue = {24},
  pages = {240602},
  numpages = {7},
  year = {2024},
  month = {Jun},
  publisher = {American Physical Society},
  doi = {10.1103/PhysRevLett.132.240602},
  url = {https://link.aps.org/doi/10.1103/PhysRevLett.132.240602}
}

@ARTICLE{9714418,
  author={Lai, Ching-Yi and Cheng, Hao-Chung},
  journal={IEEE Transactions on Information Theory}, 
  title={{Learning Quantum Circuits of Some T Gates}}, 
  year={2022},
  volume={68},
  number={6},
  pages={3951-3964},
  doi={10.1109/TIT.2022.3151760}}

@article{Lami2023,
  title = {{Nonstabilizerness via Perfect Pauli Sampling of Matrix Product States}},
  author = {Lami, Guglielmo and Collura, Mario},
  journal = {Phys. Rev. Lett.},
  volume = {131},
  issue = {18},
  pages = {180401},
  numpages = {6},
  year = {2023},
  month = {Oct},
  publisher = {American Physical Society},
  doi = {10.1103/PhysRevLett.131.180401},
  url = {https://link.aps.org/doi/10.1103/PhysRevLett.131.180401}
}

@article{PhysRevLett.133.010602,
  title = {{Unveiling the Stabilizer Group of a Matrix Product State}},
  author = {Lami, Guglielmo and Collura, Mario},
  journal = {Phys. Rev. Lett.},
  volume = {133},
  issue = {1},
  pages = {010602},
  numpages = {6},
  year = {2024},
  month = {Jul},
  publisher = {American Physical Society},
  doi = {10.1103/PhysRevLett.133.010602},
  url = {https://link.aps.org/doi/10.1103/PhysRevLett.133.010602}
}

@article{PhysRevLett.133.150604,
  title = {{Hybrid Stabilizer Matrix Product Operator}},
  author = {Mello, Antonio Francesco and Santini, Alessandro and Collura, Mario},
  journal = {Phys. Rev. Lett.},
  volume = {133},
  issue = {15},
  pages = {150604},
  numpages = {6},
  year = {2024},
  month = {Oct},
  publisher = {American Physical Society},
  doi = {10.1103/PhysRevLett.133.150604},
  url = {https://link.aps.org/doi/10.1103/PhysRevLett.133.150604}
}

@misc{mello2024clifforddressedtimedependentvariational,
      title={{Clifford Dressed Time-Dependent Variational Principle}}, 
      author={Antonio Francesco Mello and Alessandro Santini and Guglielmo Lami and Jacopo De Nardis and Mario Collura},
      year={2024},
      eprint={2407.01692},
      archivePrefix={arXiv},
      primaryClass={quant-ph},
}

@misc{bera2025nonstabilizernesssachdevyekitaevmodel,
      title={{Non-Stabilizerness of Sachdev-Ye-Kitaev Model}},
      author={Surajit Bera and Marco Schirò},
      year={2025},
      eprint={2502.01582},
      archivePrefix={arXiv},
      primaryClass={quant-ph},
}

@article{dowling2025bridgingentanglementmagicresources,
  title={Bridging Entanglement and Magic Resources within Operator Space},
  author={Dowling, Neil and Modi, Kavan and White, Gregory AL},
  journal={Physical Review Letters},
  volume={135},
  number={16},
  pages={160201},
  year={2025},
  publisher={APS}, 
  doi={10.1103/c7k1-xcwy},
}

@article{haug2023stabilizer,
  title={Stabilizer entropies and nonstabilizerness monotones},
  author={Haug, Tobias and Piroli, Lorenzo},
  journal={Quantum},
  volume={7},
  pages={1092},
  year={2023},
  publisher={Verein zur F{\"o}rderung des Open Access Publizierens in den Quantenwissenschaften},
  doi = {10.22331/q-2023-08-28-1092},
  url = {https://doi.org/10.22331/q-2023-08-28-1092},
}

@misc{smith2024nonstabilizernesskineticallyconstrainedrydbergatom,
      title={{Non-stabilizerness in kinetically-constrained Rydberg atom arrays}}, 
      author={Ryan Smith and Zlatko Papić and Andrew Hallam},
      year={2024},
      eprint={2406.14348},
      archivePrefix={arXiv},
      primaryClass={quant-ph}, 
}

@article{oliviero2022ising,
  title = {{Magic-state resource theory for the ground state of the transverse-field Ising model}},
  author = {Oliviero, Salvatore F. E. and Leone, Lorenzo and Hamma, Alioscia},
  journal = {Phys. Rev. A},
  volume = {106},
  issue = {4},
  pages = {042426},
  numpages = {6},
  year = {2022},
  month = {Oct},
  publisher = {American Physical Society},
  doi = {10.1103/PhysRevA.106.042426},
  url = {https://link.aps.org/doi/10.1103/PhysRevA.106.042426}
}

@article{Beverland2020,
  title = {{Lower bounds on the non-Clifford resources for quantum computations}},
  volume = {5},
  ISSN = {2058-9565},
  url = {http://dx.doi.org/10.1088/2058-9565/ab8963},
  DOI = {10.1088/2058-9565/ab8963},
  number = {3},
  journal = {Quantum Science and Technology},
  publisher = {IOP Publishing},
  author = {Beverland,  Michael and Campbell,  Earl and Howard,  Mark and Kliuchnikov,  Vadym},
  year = {2020},
  month = jun,
  pages = {035009}
}

@article{PRXQuantum.4.010301,
  title = {{Scalable Measures of Magic Resource for Quantum Computers}},
  author = {Haug, Tobias and Kim, M.S.},
  journal = {PRX Quantum},
  volume = {4},
  issue = {1},
  pages = {010301},
  numpages = {23},
  year = {2023},
  month = {Jan},
  publisher = {American Physical Society},
  doi = {10.1103/PRXQuantum.4.010301},
  url = {https://link.aps.org/doi/10.1103/PRXQuantum.4.010301}
}

@misc{leone2024stabilizer,
      title={Stabilizer entropies are monotones for magic-state resource theory}, 
      author={Lorenzo Leone and Lennart Bittel},
      year={2024},
      eprint={2404.11652},
      archivePrefix={arXiv},
      primaryClass={quant-ph}
}

@article{PhysRevLett.118.090501,
  title = {Application of a Resource Theory for Magic States to Fault-Tolerant Quantum Computing},
  author = {Howard, Mark and Campbell, Earl},
  journal = {Phys. Rev. Lett.},
  volume = {118},
  issue = {9},
  pages = {090501},
  numpages = {6},
  year = {2017},
  month = {Mar},
  publisher = {American Physical Society},
  doi = {10.1103/PhysRevLett.118.090501},
  url = {https://link.aps.org/doi/10.1103/PhysRevLett.118.090501}
}

@article{RevModPhys.82.277,
  title = {Colloquium: Area laws for the entanglement entropy},
  author = {Eisert, J. and Cramer, M. and Plenio, M. B.},
  journal = {Rev. Mod. Phys.},
  volume = {82},
  issue = {1},
  pages = {277--306},
  numpages = {0},
  year = {2010},
  month = {Feb},
  publisher = {American Physical Society},
  doi = {10.1103/RevModPhys.82.277},
  url = {https://link.aps.org/doi/10.1103/RevModPhys.82.277}
}

@article{PhysRevLett.71.1291,
  title = {Average entropy of a subsystem},
  author = {Page, Don N.},
  journal = {Phys. Rev. Lett.},
  volume = {71},
  issue = {9},
  pages = {1291--1294},
  numpages = {0},
  year = {1993},
  month = {Aug},
  publisher = {American Physical Society},
  doi = {10.1103/PhysRevLett.71.1291},
  url = {https://link.aps.org/doi/10.1103/PhysRevLett.71.1291}
}

@article{Gross2021,
  doi = {10.1007/s00220-021-04118-7},
  url = {https://doi.org/10.1007/s00220-021-04118-7},
  year = {2021},
  month = jun,
  publisher = {Springer Science and Business Media {LLC}},
  volume = {385},
  number = {3},
  pages = {1325--1393},
  author = {David Gross and Sepehr Nezami and Michael Walter},
  title = {{Schur{\textendash}Weyl Duality for the Clifford Group with Applications: Property Testing,  a Robust Hudson Theorem, and de Finetti Representations}},
  journal = {Communications in Mathematical Physics}
}

@article{Oliviero2022,
  title = {Measuring magic on a quantum processor},
  volume = {8},
  ISSN = {2056-6387},
  url = {http://dx.doi.org/10.1038/s41534-022-00666-5},
  DOI = {10.1038/s41534-022-00666-5},
  number = {1},
  journal = {npj Quantum Information},
  publisher = {Springer Science and Business Media LLC},
  author = {Oliviero,  Salvatore F. E. and Leone,  Lorenzo and Hamma,  Alioscia and Lloyd,  Seth},
  year = {2022},
  month = dec 
}

@misc{niroula2023phase,
      title={Phase transition in magic with random quantum circuits}, 
      author={Pradeep Niroula and Christopher David White and Qingfeng Wang and Sonika Johri and Daiwei Zhu and Christopher Monroe and Crystal Noel and Michael J. Gullans},
      year={2023},
      eprint={2304.10481},
      archivePrefix={arXiv},
      primaryClass={quant-ph}
}

@article{tarabunga2023manybody,
 title = {{Many-Body Magic Via Pauli-Markov Chains—From Criticality to Gauge Theories}},
  author = {Tarabunga, Poetri Sonya and Tirrito, Emanuele and Chanda, Titas and Dalmonte, Marcello},
  journal = {PRX Quantum},
  volume = {4},
  issue = {4},
  pages = {040317},
  numpages = {19},
  year = {2023},
  month = {Oct},
  publisher = {American Physical Society},
  doi = {10.1103/PRXQuantum.4.040317},
  url = {https://link.aps.org/doi/10.1103/PRXQuantum.4.040317}
}

@article{tarabunga2023magic,
  title = {{Magic in generalized Rokhsar-Kivelson wavefunctions}},
  volume = {8},
  ISSN = {2521-327X},
  url = {http://dx.doi.org/10.22331/q-2024-05-14-1347},
  DOI = {10.22331/q-2024-05-14-1347},
  journal = {Quantum},
  publisher = {Verein zur Forderung des Open Access Publizierens in den Quantenwissenschaften},
  author = {Tarabunga,  Poetri Sonya and Castelnovo,  Claudio},
  year = {2024},
  month = may,
  pages = {1347}
}

@misc{tarabunga2023critical,
      title={Critical behaviours of non-stabilizerness in quantum spin chains}, 
      author={Poetri Sonya Tarabunga},
      year={2023},
      eprint={2309.00676},
      archivePrefix={arXiv},
      primaryClass={quant-ph}
}

@article{haug2023quantifying,
  title = {{Quantifying nonstabilizerness of Matrix Product States}},
  author = {Haug, Tobias and Piroli, Lorenzo},
  journal = {Phys. Rev. B},
  volume = {107},
  issue = {3},
  pages = {035148},
  numpages = {10},
  year = {2023},
  month = {Jan},
  publisher = {American Physical Society},
  doi = {10.1103/PhysRevB.107.035148},
  url = {https://link.aps.org/doi/10.1103/PhysRevB.107.035148}
}

@misc{montanaro2017learning,
      title={{Learning stabilizer states by Bell sampling}}, 
      author={Ashley Montanaro},
      year={2017},
      eprint={1707.04012},
      archivePrefix={arXiv},
      primaryClass={quant-ph}
}

@article{Bluvstein2023,
  title={Logical quantum processor based on reconfigurable atom arrays},
  author={Bluvstein, Dolev and Evered, Simon J and Geim, Alexandra A and Li, Sophie H and Zhou, Hengyun and Manovitz, Tom and Ebadi, Sepehr and Cain, Madelyn and Kalinowski, Marcin and Hangleiter, Dominik and others},
  journal={Nature},
  volume={626},
  number={7997},
  pages={58--65},
  year={2024},
  publisher={Nature Publishing Group UK London}, 
  DOI = {10.1038/s41586-023-06927-3},
}

@misc{tarabunga2024nonstabilizerness,
      title={{Nonstabilizerness via Matrix Product States in the Pauli basis}}, 
      author={Poetri Sonya Tarabunga and Emanuele Tirrito and Mari Carmen Bañuls and Marcello Dalmonte},
      year={2024},
      eprint={2401.16498},
      archivePrefix={arXiv},
      primaryClass={quant-ph}
}

@article{lami2024quantum,
  title={Quantum state designs with clifford-enhanced matrix product states},
  author={Lami, Guglielmo and Haug, Tobias and De Nardis, Jacopo},
  journal={PRX Quantum},
  volume={6},
  number={1},
  pages={010345},
  year={2025},
  publisher={APS}, 
  doi={10.1103/PRXQuantum.6.010345}, 
}

@article{havlicek2023amplitude,
  doi = {10.22331/q-2023-03-02-938},
  url = {https://doi.org/10.22331/q-2023-03-02-938},
  title = {Amplitude {R}atios and {N}eural {N}etwork {Q}uantum {S}tates},
  author = {Havlicek, Vojtech},
  journal = {{Quantum}},
  issn = {2521-327X},
  publisher = {{Verein zur F{\"{o}}rderung des Open Access Publizierens in den Quantenwissenschaften}},
  volume = {7},
  pages = {938},
  month = mar,
  year = {2023}
}

@article{paviglianiti2024estimating,
  title={Estimating nonstabilizerness dynamics without simulating it},
  author={Paviglianiti, Alessio and Lami, Guglielmo and Collura, Mario and Silva, Alessandro},
  journal={PRX Quantum},
  volume={6},
  number={3},
  pages={030320},
  year={2025},
  publisher={APS}, 
  doi = {10.1103/msm2-vmg7},
}

@article{PhysRevX.7.021021,
  title = {{Quantum Entanglement in Neural Network States}},
  author = {Deng, Dong-Ling and Li, Xiaopeng and Das Sarma, S.},
  journal = {Phys. Rev. X},
  volume = {7},
  issue = {2},
  pages = {021021},
  numpages = {17},
  year = {2017},
  month = {May},
  publisher = {American Physical Society},
  doi = {10.1103/PhysRevX.7.021021},
  url = {https://link.aps.org/doi/10.1103/PhysRevX.7.021021}
}

@article{carleo2017solving,
  doi = {10.1126/science.aag2302},
  url = {https://doi.org/10.1126/science.aag2302},
  year = {2017},
  month = feb,
  publisher = {American Association for the Advancement of Science ({AAAS})},
  volume = {355},
  number = {6325},
  pages = {602--606},
  author = {Giuseppe Carleo and Matthias Troyer},
  title = {Solving the quantum many-body problem with artificial neural networks},
  journal = {Science}
}

@article{hu2013direct,
  title = {{Direct evidence for a gapless ${Z}_{2}$ spin liquid by frustrating N\'eel antiferromagnetism}},
  author = {Hu, Wen-Jun and Becca, Federico and Parola, Alberto and Sorella, Sandro},
  journal = {Phys. Rev. B},
  volume = {88},
  issue = {6},
  pages = {060402},
  numpages = {5},
  year = {2013},
  month = {Aug},
  publisher = {American Physical Society},
  doi = {10.1103/PhysRevB.88.060402},
  url = {https://link.aps.org/doi/10.1103/PhysRevB.88.060402}
}

@article{ferrari2019neural,
  title = {{Neural Gutzwiller-projected variational wave functions}},
  author = {Ferrari, Francesco and Becca, Federico and Carrasquilla, Juan},
  journal = {Phys. Rev. B},
  volume = {100},
  issue = {12},
  pages = {125131},
  numpages = {13},
  year = {2019},
  month = {Sep},
  publisher = {American Physical Society},
  doi = {10.1103/PhysRevB.100.125131},
  url = {https://link.aps.org/doi/10.1103/PhysRevB.100.125131}
}

@article{choo2019two,
  title = {Two-dimensional frustrated ${J}_{1}\text{\ensuremath{-}}{J}_{2}$ model studied with neural network quantum states},
  author = {Choo, Kenny and Neupert, Titus and Carleo, Giuseppe},
  journal = {Phys. Rev. B},
  volume = {100},
  issue = {12},
  pages = {125124},
  numpages = {7},
  year = {2019},
  month = {Sep},
  publisher = {American Physical Society},
  doi = {10.1103/PhysRevB.100.125124},
  url = {https://link.aps.org/doi/10.1103/PhysRevB.100.125124}
}

@article{liang2018solving,
  title = {Solving frustrated quantum many-particle models with convolutional neural networks},
  author = {Liang, Xiao and Liu, Wen-Yuan and Lin, Pei-Ze and Guo, Guang-Can and Zhang, Yong-Sheng and He, Lixin},
  journal = {Phys. Rev. B},
  volume = {98},
  issue = {10},
  pages = {104426},
  numpages = {6},
  year = {2018},
  month = {Sep},
  publisher = {American Physical Society},
  doi = {10.1103/PhysRevB.98.104426},
  url = {https://link.aps.org/doi/10.1103/PhysRevB.98.104426}
}

@article{szabo2020neural,
  title = {Neural network wave functions and the sign problem},
  author = {Szab\'o, Attila and Castelnovo, Claudio},
  journal = {Phys. Rev. Res.},
  volume = {2},
  issue = {3},
  pages = {033075},
  numpages = {12},
  year = {2020},
  month = {Jul},
  publisher = {American Physical Society},
  doi = {10.1103/PhysRevResearch.2.033075},
  url = {https://link.aps.org/doi/10.1103/PhysRevResearch.2.033075}
}

@article{roth2023high,
  title = {{High-accuracy variational Monte Carlo for frustrated magnets with deep neural networks}},
  author = {Roth, Christopher and Szab\'o, Attila and MacDonald, Allan H.},
  journal = {Phys. Rev. B},
  volume = {108},
  issue = {5},
  pages = {054410},
  numpages = {12},
  year = {2023},
  month = {Aug},
  publisher = {American Physical Society},
  doi = {10.1103/PhysRevB.108.054410},
  url = {https://link.aps.org/doi/10.1103/PhysRevB.108.054410}
}

@ARTICLE{li2022bridging,
  author={Li, Mingfan and Chen, Junshi and Xiao, Qian and Wang, Fei and Jiang, Qingcai and Zhao, Xuncheng and Lin, Rongfen and An, Hong and Liang, Xiao and He, Lixin},
  journal={IEEE Transactions on Parallel and Distributed Systems}, 
  title={{Bridging the Gap between Deep Learning and Frustrated Quantum Spin System for Extreme-Scale Simulations on New Generation of Sunway Supercomputer}}, 
  year={2022},
  volume={33},
  number={11},
  pages={2846-2859},
  doi={10.1109/TPDS.2022.3145163}}

@article{chen2024empowering,
  title={Empowering deep neural quantum states through efficient optimization},
  author={Chen, Ao and Heyl, Markus},
  journal={Nature Physics},
  volume={20},
  number={9},
  pages={1476--1481},
  year={2024},
  publisher={Nature Publishing Group UK London}, 
  doi={10.1038/s41567-024-02566-1}, 
}

@article{liang2023deep,
  title={Deep learning representations for quantum many-body systems on heterogeneous hardware},
  author={Liang, Xiao and Li, Mingfan and Xiao, Qian and Chen, Junshi and Yang, Chao and An, Hong and He, Lixin},
  journal={Machine Learning: Science and Technology},
  volume={4},
  number={1},
  pages={015035},
  year={2023},
  publisher={IOP Publishing}, 
  doi={10.1088/2632-2153/acc56a}
}

@article{nomura2021dirac,
  title = {{Dirac-Type Nodal Spin Liquid Revealed by Refined Quantum Many-Body Solver Using Neural-Network Wave Function, Correlation Ratio, and Level Spectroscopy}},
  author = {Nomura, Yusuke and Imada, Masatoshi},
  journal = {Phys. Rev. X},
  volume = {11},
  issue = {3},
  pages = {031034},
  numpages = {19},
  year = {2021},
  month = {Aug},
  publisher = {American Physical Society},
  doi = {10.1103/PhysRevX.11.031034},
  url = {https://link.aps.org/doi/10.1103/PhysRevX.11.031034}
}

@inproceedings{chen2022systematic, author = {Chen, Hongwei and Hendry, Douglas and Weinberg, Phillip and Feiguin, Adrian E.}, title = {{Systematic improvement of neural network quantum states using a Lanczos recursion}}, year = {2024}, isbn = {9781713871088}, publisher = {Curran Associates Inc.}, address = {Red Hook, NY, USA}, booktitle = {Proceedings of the 36th International Conference on Neural Information Processing Systems}, articleno = {544}, numpages = {14}, location = {New Orleans, LA, USA}, series = {NIPS '22}, 
}

@article{reh2023optimizing,
  title = {Optimizing design choices for neural quantum states},
  author = {Reh, Moritz and Schmitt, Markus and G\"arttner, Martin},
  journal = {Phys. Rev. B},
  volume = {107},
  issue = {19},
  pages = {195115},
  numpages = {10},
  year = {2023},
  month = {May},
  publisher = {American Physical Society},
  doi = {10.1103/PhysRevB.107.195115},
  url = {https://link.aps.org/doi/10.1103/PhysRevB.107.195115}
}

@article{wang2024variational,
  title = {Variational optimization of the amplitude of neural-network quantum many-body ground states},
  author = {Wang, Jia-Qi and Wu, Han-Qing and He, Rong-Qiang and Lu, Zhong-Yi},
  journal = {Phys. Rev. B},
  volume = {109},
  issue = {24},
  pages = {245120},
  numpages = {8},
  year = {2024},
  month = {Jun},
  publisher = {American Physical Society},
  doi = {10.1103/PhysRevB.109.245120},
  url = {https://link.aps.org/doi/10.1103/PhysRevB.109.245120}
}

@article{liang2021hybrid,
  title = {Hybrid convolutional neural network and projected entangled pair states wave functions for quantum many-particle states},
  author = {Liang, Xiao and Dong, Shao-Jun and He, Lixin},
  journal = {Phys. Rev. B},
  volume = {103},
  issue = {3},
  pages = {035138},
  numpages = {7},
  year = {2021},
  month = {Jan},
  publisher = {American Physical Society},
  doi = {10.1103/PhysRevB.103.035138},
  url = {https://link.aps.org/doi/10.1103/PhysRevB.103.035138}
}

@article{ledinauskas2023scalable,
  title={Scalable imaginary time evolution with neural network quantum states},
  author={Ledinauskas, Eimantas and Anisimovas, Egidijus},
  journal={SciPost Physics},
  volume={15},
  number={6},
  pages={229},
  year={2023}, 
  doi={10.21468/SciPostPhys.15.6.229},
}

@article{pei2021compact,
  title={{Compact neural-network quantum state representations of Jastrow and stabilizer states}},
  author={Pei, Michael Y and Clark, Stephen R},
  journal={Journal of Physics A: Mathematical and Theoretical},
  volume={54},
  number={40},
  pages={405304},
  year={2021},
  publisher={IOP Publishing}, 
  doi={10.1088/1751-8121/ac1f3d}
}

@article{lu2019efficient,
  title = {{Efficient representation of topologically ordered states with restricted Boltzmann machines}},
  author = {Lu, Sirui and Gao, Xun and Duan, L.-M.},
  journal = {Phys. Rev. B},
  volume = {99},
  issue = {15},
  pages = {155136},
  numpages = {8},
  year = {2019},
  month = {Apr},
  publisher = {American Physical Society},
  doi = {10.1103/PhysRevB.99.155136},
  url = {https://link.aps.org/doi/10.1103/PhysRevB.99.155136}
}

@article{zheng2019restricted,
  title = {{Restricted Boltzmann machines and matrix product states of one-dimensional translationally invariant stabilizer codes}},
  author = {Zheng, Yunqin and He, Huan and Regnault, Nicolas and Bernevig, B. Andrei},
  journal = {Phys. Rev. B},
  volume = {99},
  issue = {15},
  pages = {155129},
  numpages = {32},
  year = {2019},
  month = {Apr},
  publisher = {American Physical Society},
  doi = {10.1103/PhysRevB.99.155129},
  url = {https://link.aps.org/doi/10.1103/PhysRevB.99.155129}
}

@article{zhang2018efficient,
  title={An efficient algorithmic way to construct Boltzmann machine representations for arbitrary stabilizer code},
  author={Zhang, Yuan-Hang and Jia, Zhian and Wu, Yu-Chun and Guo, Guang-Can},
  journal={Entropy},
  volume={27},
  number={6},
  pages={627},
  year={2025},
  publisher={MDPI}, 
  doi={10.3390/e27060627}, 
}

@article{jia2019efficient,
  title = {Efficient machine-learning representations of a surface code with boundaries, defects, domain walls, and twists},
  author = {Jia, Zhih-Ahn and Zhang, Yuan-Hang and Wu, Yu-Chun and Kong, Liang and Guo, Guang-Can and Guo, Guo-Ping},
  journal = {Phys. Rev. A},
  volume = {99},
  issue = {1},
  pages = {012307},
  numpages = {15},
  year = {2019},
  month = {Jan},
  publisher = {American Physical Society},
  doi = {10.1103/PhysRevA.99.012307},
  url = {https://link.aps.org/doi/10.1103/PhysRevA.99.012307}
}

@article{sharir2022neural,
  title = {Neural tensor contractions and the expressive power of deep neural quantum states},
  author = {Sharir, Or and Shashua, Amnon and Carleo, Giuseppe},
  journal = {Phys. Rev. B},
  volume = {106},
  issue = {20},
  pages = {205136},
  numpages = {12},
  year = {2022},
  month = {Nov},
  publisher = {American Physical Society},
  doi = {10.1103/PhysRevB.106.205136},
  url = {https://link.aps.org/doi/10.1103/PhysRevB.106.205136}
}

@article{wu2023tensor,
  title = {From tensor-network quantum states to tensorial recurrent neural networks},
  author = {Wu, Dian and Rossi, Riccardo and Vicentini, Filippo and Carleo, Giuseppe},
  journal = {Phys. Rev. Res.},
  volume = {5},
  issue = {3},
  pages = {L032001},
  numpages = {6},
  year = {2023},
  month = {Jul},
  publisher = {American Physical Society},
  doi = {10.1103/PhysRevResearch.5.L032001},
  url = {https://link.aps.org/doi/10.1103/PhysRevResearch.5.L032001}
}

@article{white1992density,
  title = {Density matrix formulation for quantum renormalization groups},
  author = {White, Steven R.},
  journal = {Phys. Rev. Lett.},
  volume = {69},
  issue = {19},
  pages = {2863--2866},
  numpages = {0},
  year = {1992},
  month = {Nov},
  publisher = {American Physical Society},
  doi = {10.1103/PhysRevLett.69.2863},
  url = {https://link.aps.org/doi/10.1103/PhysRevLett.69.2863}
}

@article{orus2014practical,
  title={{A practical introduction to Tensor Networks: Matrix Product States and Projected Entangled Pair States}},
  author={Or{\'u}s, Rom{\'a}n},
  journal={Annals of physics},
  volume={349},
  pages={117--158},
  year={2014},
  publisher={Elsevier}, 
  doi = {10.1016/j.aop.2014.06.013},
  url = {https://doi.org/10.1016/j.aop.2014.06.013},
}

@misc{liu2024non,
      title={{Non-equilibrium quantum Monte Carlo algorithm for stabilizer Renyi entropy in spin systems}},
      author={Liu, Zejun and Clark, Bryan K},
      year={2024},
      eprint={2405.19577},
      archivePrefix={arXiv},
      primaryClass={quant-ph}
}

@article{troyer2005computational,
  title = {{Computational Complexity and Fundamental Limitations to Fermionic Quantum Monte Carlo Simulations}},
  author = {Troyer, Matthias and Wiese, Uwe-Jens},
  journal = {Phys. Rev. Lett.},
  volume = {94},
  issue = {17},
  pages = {170201},
  numpages = {4},
  year = {2005},
  month = {May},
  publisher = {American Physical Society},
  doi = {10.1103/PhysRevLett.94.170201},
  url = {https://link.aps.org/doi/10.1103/PhysRevLett.94.170201}
}

@misc{best2010simulating,
      title={Simulating quantum computers with probabilistic methods}, 
      author={M. Van den Nest},
      year={2010},
      eprint={0911.1624},
      archivePrefix={arXiv},
      primaryClass={quant-ph}
}

@article{viteritti2023transformer_1d,
  title = {{Transformer Variational Wave Functions for Frustrated Quantum Spin Systems}},
  author = {Viteritti, Luciano Loris and Rende, Riccardo and Becca, Federico},
  journal = {Phys. Rev. Lett.},
  volume = {130},
  issue = {23},
  pages = {236401},
  numpages = {6},
  year = {2023},
  month = {Jun},
  publisher = {American Physical Society},
  doi = {10.1103/PhysRevLett.130.236401},
  url = {https://link.aps.org/doi/10.1103/PhysRevLett.130.236401}
}

@misc{viteritti2023transformer_2d,
      title={{Transformer Wave Function for two dimensional frustrated magnets: emergence of a Spin-Liquid Phase in the Shastry-Sutherland Model}},
      author={Viteritti, Luciano Loris and Rende, Riccardo and Parola, Alberto and Goldt, Sebastian and Becca, Federico},
      year={2023},
      eprint={2311.16889},
      archivePrefix={arXiv},
      primaryClass={quant-ph}
}

@article{rende2024simple,
  title={A simple linear algebra identity to optimize large-scale neural network quantum states},
  author={Rende, Riccardo and Viteritti, Luciano Loris and Bardone, Lorenzo and Becca, Federico and Goldt, Sebastian},
  journal={Communications Physics},
  volume={7},
  number={1},
  pages={260},
  year={2024},
  publisher={Nature Publishing Group UK London}, 
  doi={10.1038/s42005-024-01732-4}
}

@article{sorella1998green,
  title = {{Green Function Monte Carlo with Stochastic Reconfiguration}},
  author = {Sorella, Sandro},
  journal = {Phys. Rev. Lett.},
  volume = {80},
  issue = {20},
  pages = {4558--4561},
  numpages = {0},
  year = {1998},
  month = {May},
  publisher = {American Physical Society},
  doi = {10.1103/PhysRevLett.80.4558},
  url = {https://link.aps.org/doi/10.1103/PhysRevLett.80.4558}
}

@article{sorella2005wave,
  title = {{Wave function optimization in the variational Monte Carlo method}},
  author = {Sorella, Sandro},
  journal = {Phys. Rev. B},
  volume = {71},
  issue = {24},
  pages = {241103},
  numpages = {4},
  year = {2005},
  month = {Jun},
  publisher = {American Physical Society},
  doi = {10.1103/PhysRevB.71.241103},
  url = {https://link.aps.org/doi/10.1103/PhysRevB.71.241103}
}

@article{marshall1955antiferromagnetism,
  title={Antiferromagnetism},
  author={Marshall, W},
  journal={Proceedings of the Royal Society of London. Series A. Mathematical and Physical Sciences},
  volume={232},
  number={1188},
  pages={48--68},
  year={1955},
  publisher={The Royal Society London}, 
  doi={10.1098/rspa.1955.0200}
}

@inproceedings{liu2022convnet,
  title={{A ConvNet for the 2020s}},
  author={Liu, Zhuang and Mao, Hanzi and Wu, Chao-Yuan and Feichtenhofer, Christoph and Darrell, Trevor and Xie, Saining},
  booktitle={Proceedings of the IEEE/CVF conference on computer vision and pattern recognition},
  pages={11976--11986},
  year={2022}, 
  doi={10.1109/CVPR52688.2022.01167}
}

@misc{denis2024accurate,
      title={Accurate neural quantum states for interacting lattice bosons},
      author={Denis, Zakari and Carleo, Giuseppe},
      year={2024},
      eprint={2404.07869},
      archivePrefix={arXiv},
      primaryClass={quant-ph}
}

@article{HastingsPRL2010,
  title = {{Measuring Renyi Entanglement Entropy in Quantum Monte Carlo Simulations}},
  author = {Hastings, Matthew B. and Gonz\'alez, Iv\'an and Kallin, Ann B. and Melko, Roger G.},
  journal = {Phys. Rev. Lett.},
  volume = {104},
  issue = {15},
  pages = {157201},
  numpages = {4},
  year = {2010},
  month = {Apr},
  publisher = {American Physical Society},
  doi = {10.1103/PhysRevLett.104.157201},
  url = {https://link.aps.org/doi/10.1103/PhysRevLett.104.157201}
}

@article{majumdar1969next,
  title={{On Next‐Nearest‐Neighbor Interaction in Linear Chain. I}},
  author={Majumdar, Chanchal K and Ghosh, Dipan K},
  journal={Journal of Mathematical Physics},
  volume={10},
  number={8},
  pages={1388--1398},
  year={1969},
  publisher={American Institute of Physics}, 
  doi={10.1063/1.1664978}
}

@article{haghshenas2018u,
  title = {{$U(1)$-symmetric infinite projected entangled-pair states study of the spin-1/2 square ${J}_{1}\text{\ensuremath{-}}{J}_{2}$ Heisenberg model}},
  author = {Haghshenas, R. and Sheng, D. N.},
  journal = {Phys. Rev. B},
  volume = {97},
  issue = {17},
  pages = {174408},
  numpages = {10},
  year = {2018},
  month = {May},
  publisher = {American Physical Society},
  doi = {10.1103/PhysRevB.97.174408},
  url = {https://link.aps.org/doi/10.1103/PhysRevB.97.174408}
}

@article{morita2015quantum,
  title={{Quantum Spin Liquid in Spin 1/2 J1–J2 Heisenberg Model on Square Lattice: Many-Variable Variational Monte Carlo Study Combined with Quantum-Number Projections}},
  author={Morita, Satoshi and Kaneko, Ryui and Imada, Masatoshi},
  journal={journal of the physical society of japan},
  volume={84},
  number={2},
  pages={024720},
  year={2015},
  publisher={The Physical Society of Japan}, 
  doi={10.7566/JPSJ.84.024720}
}

@article{wang2018critical,
  title = {{Critical Level Crossings and Gapless Spin Liquid in the Square-Lattice Spin-$1/2$ ${J}_{1}\ensuremath{-}{J}_{2}$ Heisenberg Antiferromagnet}},
  author = {Wang, Ling and Sandvik, Anders W.},
  journal = {Phys. Rev. Lett.},
  volume = {121},
  issue = {10},
  pages = {107202},
  numpages = {7},
  year = {2018},
  month = {Sep},
  publisher = {American Physical Society},
  doi = {10.1103/PhysRevLett.121.107202},
  url = {https://link.aps.org/doi/10.1103/PhysRevLett.121.107202}
}

@article{schmitt2020quantum,
  title = {{Quantum Many-Body Dynamics in Two Dimensions with Artificial Neural Networks}},
  author = {Schmitt, Markus and Heyl, Markus},
  journal = {Phys. Rev. Lett.},
  volume = {125},
  issue = {10},
  pages = {100503},
  numpages = {7},
  year = {2020},
  month = {Sep},
  publisher = {American Physical Society},
  doi = {10.1103/PhysRevLett.125.100503},
  url = {https://link.aps.org/doi/10.1103/PhysRevLett.125.100503}
}

@article{sinibaldi2023unbiasing,
  doi = {10.22331/q-2023-10-10-1131},
  url = {https://doi.org/10.22331/q-2023-10-10-1131},
  title = {Unbiasing time-dependent {V}ariational {M}onte {C}arlo by projected quantum evolution},
  author = {Sinibaldi, Alessandro and Giuliani, Clemens and Carleo, Giuseppe and Vicentini, Filippo},
  journal = {{Quantum}},
  issn = {2521-327X},
  publisher = {{Verein zur F{\"{o}}rderung des Open Access Publizierens in den Quantenwissenschaften}},
  volume = {7},
  pages = {1131},
  month = oct,
  year = {2023}
}

@misc{gravina2024neural,
      title={{Neural Projected Quantum Dynamics: a systematic study}},
      author={Gravina, Luca and Savona, Vincenzo and Vicentini, Filippo},
      year={2024},
      eprint={2410.10720},
      archivePrefix={arXiv},
      primaryClass={quant-ph}
}

@article{netket2,
  title={{NetKet}: {A} machine learning toolkit for many-body quantum systems},
  author={Carleo, Giuseppe and others},
  journal={SoftwareX},
  pages={100311},
  year={2019},
  publisher={Elsevier},
  doi={10.1016/j.softx.2019.100311}
}

@Article{vicentini2022netket,
	title={{NetKet 3: Machine Learning Toolbox for Many-Body Quantum Systems}},
	author={Filippo Vicentini and others},
	journal={SciPost Phys. Codebases},
	pages={7},
	year={2022},
	publisher={SciPost},
	doi={10.21468/SciPostPhysCodeb.7},
	url={https://scipost.org/10.21468/SciPostPhysCodeb.7},
}

@article{gidney2021stim,
  doi = {10.22331/q-2021-07-06-497},
  url = {https://doi.org/10.22331/q-2021-07-06-497},
  title = {Stim: a fast stabilizer circuit simulator},
  author = {Gidney, Craig},
  journal = {{Quantum}},
  issn = {2521-327X},
  publisher = {{Verein zur F{\"{o}}rderung des Open Access Publizierens in den Quantenwissenschaften}},
  volume = {5},
  pages = {497},
  month = jul,
  year = {2021}
}

@misc{verstraete2004renormalization,
      title={{Renormalization algorithms for quantum-many body systems in two and higher dimensions}},
      author={Verstraete, Frank and Cirac, J Ignacio},
      year={2004},
      eprint={cond-mat/0407066},
      archivePrefix={arXiv},
}

@misc{sinibaldi2024time,
  title={{Time-dependent Neural Galerkin Method for Quantum Dynamics}},
  author={Sinibaldi, Alessandro and Hendry, Douglas and Vicentini, Filippo and Carleo, Giuseppe},
  year={2024},
  eprint={2412.11778},
  archivePrefix={arXiv},
  primaryClass={quant-ph}
}

@article{viteritti2022accuracy,
  title={{Accuracy of restricted Boltzmann machines for the one-dimensional ${J}_1$-${J}_2$ Heisenberg model}},
  author={Viteritti, Luciano Loris and Ferrari, Francesco and Becca, Federico},
  journal={SciPost Physics},
  volume={12},
  number={5},
  pages={166},
  year={2022}, 
  doi = {10.21468/SciPostPhys.12.5.166}
}

@article{eggert1996numerical,
  title = {Numerical evidence for multiplicative logarithmic corrections from marginal operators},
  author = {Eggert, Sebastian},
  journal = {Phys. Rev. B},
  volume = {54},
  issue = {14},
  pages = {R9612--R9615},
  numpages = {0},
  year = {1996},
  month = {Oct},
  publisher = {American Physical Society},
  doi = {10.1103/PhysRevB.54.R9612},
  url = {https://link.aps.org/doi/10.1103/PhysRevB.54.R9612}
}

@article{white1996dimerization,
  title = {{Dimerization and incommensurate spiral spin correlations in the zigzag spin chain: Analogies to the Kondo lattice}},
  author = {White, Steven R. and Affleck, Ian},
  journal = {Phys. Rev. B},
  volume = {54},
  issue = {14},
  pages = {9862--9869},
  numpages = {0},
  year = {1996},
  month = {Oct},
  publisher = {American Physical Society},
  doi = {10.1103/PhysRevB.54.9862},
  url = {https://link.aps.org/doi/10.1103/PhysRevB.54.9862}
}

@misc{denis2023comment,
  title={{Comment on "Can Neural Quantum States Learn Volume-Law Ground States?"}},
  author={Denis, Zakari and Sinibaldi, Alessandro and Carleo, Giuseppe},
  year={2023},
  eprint={2309.11534},
  archivePrefix={arXiv},
  primaryClass={quant-ph}
}

@misc{falcao2024non,
      title={{Non-stabilizerness in U (1) lattice gauge theory}},
      author={Falc{\~a}o, Pedro R Nic{\'a}cio and Tarabunga, Poetri Sonya and Frau, Martina and Tirrito, Emanuele and Zakrzewski, Jakub and Dalmonte, Marcello},
      year={2024},
      eprint={2409.01789},
      archivePrefix={arXiv},
      primaryClass={quant-ph},
    
}

@article{neal2001annealed,
  title={Annealed importance sampling},
  author={Neal, Radford M},
  journal={Statistics and computing},
  volume={11},
  pages={125--139},
  year={2001},
  publisher={Springer},
  doi={10.1023/A:1008923215028}
}

@misc{collura2025quantummagicfermionicgaussian,
      title={{The quantum magic of fermionic Gaussian states}},
      author={Mario Collura and Jacopo De Nardis and Vincenzo Alba and Guglielmo Lami},
      year={2025},
      eprint={2412.05367},
      archivePrefix={arXiv},
      primaryClass={quant-ph},
}

@article{mbeng2024quantum,
  title={The quantum Ising chain for beginners},
  author={Mbeng, Glen Bigan and Russomanno, Angelo and Santoro, Giuseppe E},
  journal={SciPost Physics Lecture Notes},
  pages={082},
  year={2024}, 
  doi={10.21468/SciPost.Report.2368}, 
}

@article{bengio2014representationlearningreviewnew,
  title={Representation learning: A review and new perspectives},
  author={Bengio, Yoshua and Courville, Aaron and Vincent, Pascal},
  journal={IEEE transactions on pattern analysis and machine intelligence},
  volume={35},
  number={8},
  pages={1798--1828},
  year={2013},
  publisher={IEEE}, 
  doi={10.1109/TPAMI.2013.50}, 
}

@article{spriggs1998quantum,
  title={Quantum resources of quantum and classical variational methods},
  author={Spriggs, Thomas and Ahmadi, Arash and Chen, Bokai and Greplova, Eliska},
  journal={Machine Learning: Science and Technology},
  volume={6},
  number={1},
  pages={015042},
  year={2025},
  publisher={IOP Publishing}, 
  doi={10.1088/2632-2153/adaca2}, 
}

@article{odavic2023complexity,
  title={Complexity of frustration: A new source of non-local non-stabilizerness},
  author={Odavi{\'c}, Jovan and Haug, Tobias and Torre, Gianpaolo and Hamma, Alioscia and Franchini, Fabio and Giampaolo, Salvatore Marco},
  journal={SciPost physics},
  volume={15},
  number={4},
  pages={131},
  year={2023}, 
  doi={10.21468/SciPostPhys.15.4.131},
}

@article{PhysRevB.109.L161103,
  title = {Absence of spin liquid phase in the ${J}_{1}\ensuremath{-}{J}_{2}$ Heisenberg model on the square lattice},
  author = {Qian, Xiangjian and Qin, Mingpu},
  journal = {Phys. Rev. B},
  volume = {109},
  issue = {16},
  pages = {L161103},
  numpages = {7},
  year = {2024},
  month = {Apr},
  publisher = {American Physical Society},
  doi = {10.1103/PhysRevB.109.L161103},
  url = {https://link.aps.org/doi/10.1103/PhysRevB.109.L161103}
}

@article{PhysRevLett.113.027201,
  title = {Plaquette Ordered Phase and Quantum Phase Diagram in the Spin-$\frac{1}{2}$ ${J}_{1}\text{\ensuremath{-}}{J}_{2}$ Square Heisenberg Model},
  author = {Gong, Shou-Shu and Zhu, Wei and Sheng, D. N. and Motrunich, Olexei I. and Fisher, Matthew P. A.},
  journal = {Phys. Rev. Lett.},
  volume = {113},
  issue = {2},
  pages = {027201},
  numpages = {5},
  year = {2014},
  month = {Jul},
  publisher = {American Physical Society},
  doi = {10.1103/PhysRevLett.113.027201},
  url = {https://link.aps.org/doi/10.1103/PhysRevLett.113.027201}
}

@misc{suppmat,
      title={{See Supplemental Material for additional details on the two Monte Carlo estimators for the SRE $M_2$ and their benchmark against MPS, the annealed importance sampling scheme, the optimization of the Deep ViT ansatz, and the variational error on the SRE density.}}, 
      author = { },
}

@software{repo,
  author       = {Sinibaldi, Alessandro},
  title        = {cqsl/nqsmagic: v0.0.1},
  month        = aug,
  year         = 2025,
  publisher    = {Zenodo},
  version      = {v0.0.1},
  doi          = {10.5281/zenodo.16941784},
  url          = {https://doi.org/10.5281/zenodo.16941784}
}

@article{nys2024ab,
  title={Ab-initio variational wave functions for the time-dependent many-electron Schr{\"o}dinger equation},
  author={Nys, Jannes and Pescia, Gabriel and Sinibaldi, Alessandro and Carleo, Giuseppe},
  journal={Nature Communications},
  volume={15},
  number={1},
  pages={9404},
  year={2024},
  doi={10.1038/s41467-024-53672-w},
  publisher={Nature Publishing Group UK London}
}

\begin{widetext}

\newpage 

\section*{SUPPLEMENTAL MATERIAL}

\subsection{Replicated estimator for $M_2$ \label{sec:replicated_estimator}}

As shown in~\cite{tarabunga2023magic}, for a generic many-qubit state $\ket{\Psi} = \sum_\sigma \frac{\Psi(\sigma)}{\sqrt{Z}} \ket{\sigma} \in \mathcal{H}$ with $Z = \sum_{\sigma} |\Psi(\sigma)|^2$, it is true that:
\begin{equation}
\label{eq:poetry_formula}
\begin{split}
    \exp(-M_2) = \sum_{\substack{\sigma^{(1)}, \sigma^{(2)}, \\ \sigma^{(3)}, \sigma^{(4)}}} \frac{1}{Z^4}\bigg[&\Psi(\sigma^{(1)})\Psi(\sigma^{(2)})\Psi(\sigma^{(3)})  \Psi^*(\sigma^{(4)}) \Psi^*(\sigma^{(2)} \odot \sigma^{(3)} \odot \sigma^{(4)}) \Psi^*(\sigma^{(1)} \odot \sigma^{(3)} \odot \sigma^{(4)}) \\
    &\Psi^*(\sigma^{(1)} \odot \sigma^{(2)} \odot \sigma^{(4)}) \Psi(\sigma^{(1)} \odot \sigma^{(2)} \odot \sigma^{(3)})\bigg], 
\end{split}
\end{equation}
where $\sigma^{(i)}$ for $i=1, \ldots, 4$ are many-qubit configurations and $\odot$ denotes the Hadamard product between qubit strings. 
Here, we consider the local spin basis $\{-1, 1\}$ instead of $\{0, 1\}$.
The idea to the replicated estimator for $M_2$ is indexing the Pauli strings with two $N$-bit indexes and expanding the fourth power appearing in $M_2$. 
Then, it is possible to carry out the sum over one $N$-bit index explicitly and derive a constraint that makes the spins in the four configurations interact.

By introducing the state $\ket{\Phi} = \ket{\Psi^*, \Psi^*, \Psi^*, \Psi}$ belonging to $\bigotimes_{i=1}^4 \mathcal{H}$ and the operator on the same replicated space acting as $\hat U|\boldsymbol{\eta}\rangle = \hat U|\sigma^{(1)}, \sigma^{(2)}, \sigma^{(3)}, \sigma^{(4)}\rangle	=	|\sigma^{(2)}\odot\sigma^{(3)}\odot\sigma^{(4)}, \sigma^{(1)}\odot\sigma^{(3)}\odot\sigma^{(4)},\sigma^{(1)}\odot\sigma^{(2)}\odot\sigma^{(4)},\sigma^{(1)}\odot\sigma^{(2)}\odot\sigma^{(3)}\rangle$,~\cref{eq:poetry_formula} can be written as:
\begin{equation}
\label{eq:poetry_as_expected_value}
    \exp(-M_2) = \frac{\bra{\Phi} \hat U \ket{\Phi}}{\bra{\Phi}\ket{\Phi}} = \sum_{\boldsymbol{\eta}} \frac{|\Phi(\boldsymbol{\eta})|^2}{\bra{\Phi}\ket{\Phi}}\bigg[\frac{\bra{\boldsymbol{\eta}} \hat U \ket{\Phi}}{\bra{\boldsymbol{\eta}}\ket{\Phi}}\bigg] = \mathbb{E}_{\boldsymbol{\eta} \sim |\Phi(\boldsymbol{\eta})|^2}\bigg[\frac{\bra{\boldsymbol{\eta}} \hat U \ket{\Phi}}{\bra{\boldsymbol{\eta}}\ket{\Phi}}\bigg]. 
\end{equation} 

The expectation value in~\cref{eq:poetry_as_expected_value} can be estimated by stochastic sampling from $|\Phi(\boldsymbol{\eta})|^2 / \bra{\Phi} \ket{\Phi}$.
We observe that the operator $\hat U$ performs a change of basis in the replicated Hilbert space. 
In particular, it is involutory since $\hat U = \hat{U}^{-1}$. 

\subsection{Bell basis estimator for $M_2$\label{sec:bell_basis_estimator}}
It is possible to compute $M_2$ from the definition by leveraging the proportionality between the expectation value of a Pauli string and the amplitude over a Bell state between the original system and a replica of it, where the state in the replica is conjugated~\cite{montanaro2017learning,Gross2021,9714418,PRXQuantum.4.010301,PhysRevLett.132.240602}. 
More precisely, we have:
\begin{equation}
\label{eq:pauli_string_bell}
   \frac{|\langle \Psi|\hat P|\Psi \rangle|}{\langle \Psi|\Psi\rangle} =
    \frac{ |\braket{\boldsymbol{B}}{\Psi,\Psi^*}|}{\braket{\boldsymbol{B}_0}{\Psi,\Psi^*}},
\end{equation}
where $\hat P = \hat\sigma^{\mu_1}\hat\sigma^{\mu_2}\cdots\hat\sigma^{\mu_N}$ and $\ket{\boldsymbol{B}} = \ket*{\phi^{\mu_1},\phi^{\mu_2},\dots,\phi^{\mu_N}}$, with $\{ \hat\sigma^{\mu} \}_{\mu=0}^3$ the Pauli matrices,  $\{ \ket*{\phi^{\mu}} \}_{\mu=0}^3$ the Bell states and $\mu_j\in\{0,1,2,3\}$ for $j=1, \ldots, N$. We define $\ket{\boldsymbol{B}_0} = \ket*{\phi^{0},\phi^{0},\dots,\phi^{0}}$. 
The doubled state $\ket{\Psi,\Psi^*}$ belongs to $\mathcal{H} \otimes \mathcal{H}$ and the Bell pairs are between a qubit and its correspondent copy in the replica system.

Since in Variational Monte Carlo (VMC) the sampling is typically performed in the computational basis, we need to go back to it by applying the Clifford transformation $\hat C_{i i^\prime} = \text{CNOT}_{i i^\prime} \text{H}_{i}$ on each qubit $i$ and its copy $i^\prime$, where $\text{CNOT}_{i i^\prime} $ is the CNOT gate with control $i$ and target $i^\prime$, while $ \text{H}_{i}$ is the Hadamard gate on $i$. 
Indeed, this transformation gives 
$\ket*{\phi^0}_{i i^\prime}  =  \hat C_{i i^\prime} \ket{00}_{i i^\prime}$,
$\ket*{\phi^1}_{i i^\prime}  =  \hat C_{i i^\prime} \ket{01}_{i i^\prime}$,
$\ket*{\phi^2}_{i i^\prime}  =  \hat C_{i i^\prime} \ket{10}_{i i^\prime}$ and
$\ket*{\phi^3}_{i i^\prime}  =  \hat C_{i i^\prime} \ket{11}_{i i^\prime}$.
Therefore, considering an arbitrary basis state $\ket{\boldsymbol{\nu}} \in \mathcal{H} \otimes \mathcal{H}$, we obtain:
\begin{equation}
\label{eq:bell_computational}
\braket{\boldsymbol{B}}{\Psi,\Psi^*}  =   \braket{\boldsymbol{\nu}}{\Gamma},
\end{equation}
with the transformed state $\ket{\Gamma} = \hat C^{\dag}\ket{\Psi,\Psi^*}$
where $\hat C = \prod_{i, i^\prime} \hat C_{ii^\prime}$ and the product runs over each couple composed by a qubit and its corresponding copy. 
In particular, $\ket{\boldsymbol{B}_0}$ is mapped onto $\ket{\boldsymbol{\nu}_0} = |0, 0, \ldots, 0\rangle$.
In general, it is impractical to apply $\hat C^{\dag}$ on $\ket{\Psi,\Psi^*}$ when $\ket{\Psi}$ is an arbitrary quantum state. 
However, if $\ket{\Psi}$ is the ground state of a $N$-qubit Hamiltonian $\hat{H}$, namely $\hat H\ket{\Psi} = E \ket{\Psi}$ with $E$ the ground energy, then $\ket{\Gamma}$ can be found as the ground state of the transformed Hamiltonian $\mathbb{\hat H} = \hat C^{\dag}(\hat H \otimes \hat I + \hat I \otimes \hat H^*)\hat C$ with energy $2 E$. Notably, $\mathbb{\hat{H}}$ can be easily obtained through the stabilizer formalism~\cite{gottesman1997stabilizer,gottesman1998heisenberg,aaronson2004improved}.

Once we have $\ket{\Gamma}$, then: 
\begin{equation}
\label{eq:bell_as_expected_value}
\exp(-M_2) = 
    \frac{1}{2^{N}} \sum_{\hat P \in \mathcal{P}_N} 
    \frac{\langle \Psi|\hat P|\Psi \rangle^4}{\langle \Psi|\Psi\rangle^4}=
    \frac{1}{2^{N}}
    \sum_{\boldsymbol{\nu}} 
    \frac{|\braket{\boldsymbol{\nu}}{\Gamma}|^4}{\braket{\boldsymbol{\nu}_0}{\Gamma}^4}=
    \sum_{\boldsymbol{\nu}} 
    \frac{|\braket{\boldsymbol{\nu}}{\Gamma}|^2}{\bra{\Gamma}\ket{\Gamma}}\bigg[\frac{|\braket{\boldsymbol{\nu}}{\Gamma}|^2}{\braket{\boldsymbol{\nu}_0}{\Gamma}^2}\bigg] = \mathbb{E}_{\boldsymbol{\nu} \sim |\Gamma(\boldsymbol{\nu})|^2}\bigg[\frac{|\Gamma(\boldsymbol{\nu})|^2}{\Gamma(\boldsymbol{\nu}_0)^2}\bigg],
\end{equation}
where in the second to last passage we have used that
$\braket{\Gamma}{\Gamma} = 2^N\braket{\boldsymbol{\nu}_0}{\Gamma}^2$, which comes from the normalization of the distribution $\Xi_{\hat P}$ in the space of Pauli strings. 
The last expectation value in~\cref{eq:bell_as_expected_value} is then estimated by Monte Carlo sampling from the distribution $|\Gamma(\boldsymbol{\nu})|^2 / \bra{\Gamma} \ket{\Gamma}$.

\subsection{Statistical errors of the estimators}
Both the replicated and the Bell basis estimator for the $M_2$ can be expressed as: 
\begin{equation}
\label{eq:general_form}
    M_2 = - \log \mathbb{E}_{P(x)}[l(x)],  
\end{equation}
where $P(x)$ is a probability distribution and $l(x)$ is a suitable stochastic estimator. 
The statistical error associated to $M_2$ computed as in~\cref{eq:general_form} corresponds to: 
\begin{equation}
    \Delta M_2 = \frac{1}{\mathbb{E}_{P(x)}[l(x)]}\sqrt{\frac{\text{Var}_{P(x)}[l(x)]}{N_s}} = \frac{1}{e^{-M_2}}\sqrt{\frac{\mathbb{E}_{P(x)}[|l(x)|^2] - e^{-2 M_2}}{N_s}} , 
\end{equation}
where $\text{Var}_{P(x)}[l(x)]$ indicates the variance of $l(x)$ over the distribution $P(x)$ and $N_s$ is the number of samples used in the empirical estimations of the statistical averages.

For the replicated estimator, $P(x) = \frac{|\Phi(\boldsymbol{\eta})|^2}{\bra{\Phi}\ket{\Phi}}$ and $l(x) = \frac{\bra{\boldsymbol{\eta}} \hat{U} \ket{\Phi}}{\bra{\boldsymbol{\eta}}\ket{\Phi}}$ where $|\boldsymbol{\eta}\rangle \in  \bigotimes_{i=1}^4 \mathcal{H}$, with $\mathcal{H}$ indicating the Hilbert space.
Therefore, we have: 
\begin{equation}
    \mathbb{E}_{P(x)}[|l(x)|^2] = \sum_{\boldsymbol{\eta}} \frac{|\Phi(\boldsymbol{\eta})|^2}{\bra{\Phi}\ket{\Phi}} \bigg| \frac{\bra{\boldsymbol{\eta}} \hat{U} \ket{\Phi}}{\bra{\boldsymbol{\eta}}\ket{\Phi}}\bigg|^2 =  \sum_{\boldsymbol{\eta}} \frac{|\bra{\boldsymbol{\eta}} \hat{U} \ket{\Phi}|^2}{\bra{\Phi}\ket{\Phi}} = 1, 
\end{equation}
since $\hat{U}$ performs a change of basis in the Hilbert space and is unitary.

In the case of the replicated estimator, we thus obtain the statistical error: 
\begin{equation}
\label{eq:replicated}
    \Delta M_2^{\text{repl.}} = \sqrt{\frac{1 - e^{-2 M_2}}{N_s e^{-2 M_2}}} = \sqrt{\frac{e^{2 M_2} - 1}{N_s}}.
\end{equation}

For the Bell basis estimator, instead, $P(x) = \frac{|\Gamma(\boldsymbol{\nu})|^2}{\bra{\Gamma}\ket{\Gamma}}$ and $l(x) = \frac{|\Gamma(\boldsymbol{\nu})|^2}{\Gamma(\boldsymbol{\nu}_0)^2}$ where $|\boldsymbol{\nu}\rangle \in  \mathcal{H} \bigotimes \mathcal{H}$.
Therefore we get: 
\begin{align}
\label{eq:bell}
    \mathbb{E}_{P(x)}[|l(x)|^2] = \sum_{\boldsymbol{\nu}} \frac{|\Gamma(\boldsymbol{\nu})|^2}{\bra{\Gamma}\ket{\Gamma}} \frac{|\Gamma(\boldsymbol{\nu})|^4}{|\Gamma(\boldsymbol{\nu}_0)|^4} = \frac{1}{2^N} \sum_{\boldsymbol{\nu}} \frac{|\Gamma(\boldsymbol{\nu})|^6}{|\Gamma(\boldsymbol{\nu}_0)|^6} = \frac{1}{2^{N}} \sum_{\hat P \in \mathcal{P}_N} \frac{\langle \Psi|\hat P|\Psi \rangle^6}{\langle \Psi|\Psi\rangle^6} = e^{-2 M_3},
\end{align}
where in the second passage we employ the identity $\braket{\Gamma}{\Gamma} = 2^N\Gamma(\nu_0)^2$ already used in~\cref{eq:bell_as_expected_value}.

Finally, the error results: 
\begin{equation}
\label{eq:error_bell}
    \Delta M_2^{\text{Bell}} = \sqrt{\frac{e^{-2 M_3} - e^{-2 M_2}}{N_s e^{-2 M_2}}} = \sqrt{\frac{e^{2 (M_2 - M_3)} - 1}{N_s}}.
\end{equation}

\subsection{Annealed importance sampling for the replicated estimator\label{sec:annealed_sampling}}
The stochastic estimate in the replicated estimator is in general affected by large statistical fluctuations.
This is caused by the fact that the transformation $\hat{U}$ remarkably changes a configuration $\ket{\boldsymbol{\eta}}$ on which it is applied, and so the amplitudes $\bra{\boldsymbol{\eta}} \ket{\Phi}$ and $\bra{\boldsymbol{\eta}} \hat{U} \ket{\Phi}$ can be very different from each other. 
In this scenario, there can be rare configurations, namely for which $|\Phi(\boldsymbol{\eta})|^2/\bra{\Phi}\ket{\Phi} \ll 1$, where the estimator $\bra{\boldsymbol{\eta}} \hat{U} \ket{\Phi} / \bra{\boldsymbol{\eta}} \ket{\Phi}$ assume large values. 
These outliers with big contributions significantly skew the statistics. 
On the other hand, it can happen that for more frequent configurations the corresponding values of the estimator are negligible.
To overcome this problem, one can introduce a strategy inspired by the annealed importance sampling scheme~\cite{neal2001annealed}. 
The idea is bridging the gap between the states $\bra{\boldsymbol{\eta}} \ket{\Phi}$ and $\bra{\boldsymbol{\eta}} \hat{U} \ket{\Phi}$ by introducing a family of intermediate states which interpolate among the two. 
In particular, we can take a set of $n+1$ real numbers $\beta_0=0 < \beta_1 < \beta_2 < \ldots < \beta_{n-1} < \beta_n = 1$ and define the corresponding set of states $\bra*{\boldsymbol{\eta}}\ket*{\Phi_{\beta_i}} = \beta_i \bra{\boldsymbol{\eta}} \hat{U} \ket{\Phi} + (1 - \beta_i) \bra{\boldsymbol{\eta}}\ket{\Phi}$ for $i=0, \ldots, n$.
We note that $\bra{\boldsymbol{\eta}}\ket{\Phi_{\beta_0}} = \bra{\boldsymbol{\eta}}\ket{\Phi}$ and $\bra{\boldsymbol{\eta}}\ket{\Phi_{\beta_n}}  = \bra{\boldsymbol{\eta}} \hat{U}\ket{\Phi}$.
Therefore, we can write:
\begin{equation}
\begin{split}
\label{eq:annealed_sampling}
    \frac{\bra{\Phi} \hat{U} \ket{\Phi}}{\bra{\Phi} \ket{\Phi}} &= \frac{\sum_{\boldsymbol{\eta}}\bra{\Phi} \ket{\boldsymbol{\eta}} \bra{\boldsymbol{\eta}} \hat{U} \ket{\Phi}}{\sum_{\boldsymbol{\eta}} \bra*{\Phi} \ket*{\boldsymbol{\eta}} \bra*{\boldsymbol{\eta}} \ket*{\Phi}} = \frac{\sum_{\boldsymbol{\eta}}\bra{\Phi}\ket{\boldsymbol{\eta}}\bra{\boldsymbol{\eta}} \hat{U} \ket{\Phi}}{\sum_{\boldsymbol{\eta}}|\bra{\Phi}\ket{\boldsymbol{\eta}} \bra*{\boldsymbol{\eta}}\ket*{\Phi_{\beta_{n-1}}}|} 
    \prod_{i=0}^{n-2} \frac{\sum_{\boldsymbol{\eta}}|\bra{\Phi}\ket{\boldsymbol{\eta}} \bra*{\boldsymbol{\eta}}\ket*{\Phi_{\beta_{i+1}}}|}{\sum_{\boldsymbol{\eta}}|\bra{\Phi}\ket{\boldsymbol{\eta}} \bra{\boldsymbol{\eta}}\ket{\Phi_{\beta_{i}}}|} = \\     
    &= \mathbb{E}_{\boldsymbol{\eta} \sim P_{n-1}(\boldsymbol{\eta})}\bigg[\frac{\bra{\Phi}\ket{\boldsymbol{\eta}}\bra{\boldsymbol{\eta}} \hat{U} \ket{\Phi}}{P_{n-1}(\boldsymbol{\eta})}\bigg] \prod_{i=0}^{n-2}  \mathbb{E}_{\boldsymbol{\eta} \sim P_{i}(\boldsymbol{\eta})}\bigg[\bigg| \frac{\bra*{\boldsymbol{\eta}}\ket*{\Phi_{\beta_{i+1}}}}{{\bra{\boldsymbol{\eta}}\ket{\Phi_{\beta_{i}}}}}\bigg| \bigg], 
\end{split}
\end{equation}
where $P_i(\boldsymbol{\eta}) = |\bra{\Phi}\ket{\boldsymbol{\eta}} \bra{\boldsymbol{\eta}}\ket{\Phi_{\beta_{i}}}|$.
By choosing sufficiently dense $\beta$ values in $[0, 1]$, the estimator of each average in~\cref{eq:annealed_sampling} remains $O(1)$ and exhibits a smooth profile across the samples.
We are able to remove the outlier problem at the price of performing multiple samplings from different distributions. 
We note that all the averages in~\cref{eq:annealed_sampling} except the first one have a positive-definite estimator.

\subsection{Benchmark on the 1D Transverse Field Ising model\label{sec:benchmark}}
We benchmark the two Monte Carlo estimators for the non-stabilizerness by measuring the magic content in the ground state of the 1D Transverse Field Ising (TFI)~\cite{mbeng2024quantum} model, whose Hamiltonian for $N$ spins-$\frac{1}{2}$ is: 
\begin{equation}
\label{eq:tfim}
    \hat{H} = -J
    \sum_{\langle i, j \rangle} \hat{\sigma}^z_{i} \hat{\sigma}^z_{j} - h\sum_{i} \hat{\sigma}^x_{i}, 
\end{equation}
where $\hat{\sigma}_i^{z}$, $\hat{\sigma}_i^{x}$ denote the $z$-$x$ Pauli matrices on the site $i$, $J$ is the coupling strength, $h$ is the intensity of the transverse magnetic field and periodic boundary conditions are assumed. 
The model is exactly solvable via the Jordan-Wigner mapping to free fermions, and displays a quantum phase transition at $h/J = 1$.
The non-stabilizerness in the 1D TFI model has been extensively studied in previous works mostly with TN methods~\cite{oliviero2022ising,Lami2023,tarabunga2023manybody,haug2023quantifying,tarabunga2024nonstabilizerness}.
Remarkably, it has been observed that the $M_2$ shows a peak in correspondence of the phase transition.

Using both the Monte Carlo estimators, we compute the SRE $M_2$ on the ground state of~\cref{eq:tfim} for many values of $h/J$ in systems with different sizes. 
The results are compared to reference calculations performed via perfect sampling with MPS~\cite{Lami2023}.
Both the physical ground state $\ket{\Psi}$ for the replicated estimator and the doubled entangled state $\ket{\Gamma}$ for the Bell basis estimator are approximated using VMC with NQS ansätze.
For the ground state $\ket{\Psi}$,  a Restricted Boltzmann Machine (RBM)~\cite{carleo2017solving} with hidden unit density $\alpha = 4$ is employed.
For the doubled state $\ket{\Gamma}$, the RBM alone is not enough expressive to capture the strong correlations introduced by the Clifford transformations on the doubled system. 
Therefore, we resort to an NQS model composed by an RBM acting on the hidden representation produced by a ConvNext architecture~\cite{liu2022convnet}, a recently introduced convolutional neural network, in the same \emph{representation learning}~\cite{bengio2014representationlearningreviewnew} fashion as done in~\cite{viteritti2023transformer_2d,denis2024accurate}. 
In the VMC for the state $\ket{\Gamma}$, we found that shifting the variational wave function with respect to its value in $\ket{\boldsymbol{\nu}_0}$, namely redefining it as $\Psi_{\theta}(\boldsymbol{\nu}) \equiv \exp(\log \Psi_{\theta}(\boldsymbol{\nu}) - \log\Psi_{\theta}(\boldsymbol{\nu}_0))$, led to more accurate results.
This improvement is likely related to the special role of the configuration $\ket{\boldsymbol{\nu}_0}$ in the Bell basis estimator.

The VMC for both the networks $\ket{\Psi_{\theta}}$ is performed using the Stochastic Reconfiguration~\cite{sorella1998green,sorella2005wave} method, where the update for the parameters $\theta$ to minimize the energy is given by:
\begin{equation}
\label{eq:SR}
        \delta \theta = \tau (S + \lambda I_{N_p})^{-1} f.
\end{equation}

In the previous equation, $S \in \mathbb{R}^{N_p \times N_p}$, where $N_p$ is the number of parameters, with elements $S_{\alpha \beta} = \Re[\mathbb{E}_{\sigma \sim |\Psi_{\theta}(\sigma)|^2}[\Bar{O}^*_{\alpha}(\sigma) \Bar{O}_{\beta}(\sigma)]]$, where $\Bar{O}_{\alpha}(\sigma) = O_{\alpha}(\sigma) - \mathbb{E}_{\sigma \sim |\Psi_{\theta}(\sigma)|^2}[O_{\alpha}(\sigma)]$ and $O_{\alpha}(\sigma)  = \partial_{\theta_\alpha} \log \Psi_{\theta}(\sigma)$ is the Jacobian.
Similarly, $f \in \mathbb{R}^{N_p}$ with elements $f_{\beta} = -2 \Re[\mathbb{E}_{\sigma \sim |\Psi_{\theta}(\sigma)|^2}[\Bar{E}_{\text{loc}}(\sigma)\Bar{O}^*_{\beta}(\sigma)]]$, where $\Bar{E}_{\text{loc}}(\sigma) = E_{\text{loc}}(\sigma) - \mathbb{E}_{\sigma \sim |\Psi_{\theta}(\sigma)|^2}[E_{\text{loc}}(\sigma)]$ and $E_{\text{loc}}(\sigma) = \bra{\sigma} \hat{H} \ket{\Psi_{\theta}} / \bra{\sigma} \ket{\Psi_{\theta}}$ is the local energy. 
In~\cref{eq:SR} $\tau$ is the learning rate, $I_{N_p}$ indicates the identity in $\mathbb{R}^{N_p \times N_p}$ and $\lambda$ is the regularization strength to ensure the invertibility of $S$. 
All the previous statistical averages are estimated through Monte Carlo sampling from $|\Psi_{\theta}(\sigma)|^2 / \bra{\Psi_{\theta}} \ket{\Psi_{\theta}}$.
In the preceding formula, $\sigma$ indicates an arbitrary configuration of the physical Hilbert space for the replicated estimator and of the doubled Hilbert space for the Bell basis estimator.
In our simulations, we use a number of samples $N_s = 8192$ and we set $\tau = 10^{-3}$ and $\lambda = 10^{-4}$.

While a simple transition rule consisting of a single-spin flip is effective in the VMC for the ground state $\ket{\Psi}$, the sampling for the doubled state $\ket{\Gamma}$ requires more attention. 
Indeed, thanks to the $\mathbb{Z}_2$ symmetry of the TFI model, the expectation value of many Pauli strings is exactly zero on the ground state. 
In the mapping from Pauli strings to Bell states, this means that several amplitudes of the state $\ket{\Gamma}$ are vanishing. 
To efficiently approximate $\ket{\Gamma}$ with VMC, it is imperative to design an ergodic sampling scheme that explores only the subspace of non-zero amplitudes. 
In~\cite{tarabunga2023manybody}, the authors perform Markov chain Monte Carlo in the space of Pauli strings and, to preserve $\mathbb{Z}_2$, they choose a transition rule consisting in applying either $\hat \sigma^x_i$ or $\hat \sigma^z_i \hat \sigma^z_j$ for randomly chosen sites $i$, $j$. 
Up to phases, applying $\hat \sigma^x$ produces the transformations $\hat I \rightarrow \hat \sigma^x$, $\hat \sigma^x \rightarrow \hat I$, $\hat \sigma^y \rightarrow \hat \sigma^z$ and $\hat \sigma^z \rightarrow \hat \sigma^y$, while applying $\hat \sigma^z$ generates $\hat I \rightarrow \hat \sigma^z$, $\hat \sigma^x \rightarrow \hat \sigma^y$, $\hat \sigma^y \rightarrow \hat \sigma^x$ and $\hat \sigma^z \rightarrow \hat I$. 
If according to~\cref{eq:pauli_string_bell,eq:bell_computational} we make the identifications $\hat I \Leftrightarrow \ket{00}$, $\hat \sigma^x \Leftrightarrow \ket{01}$, $\hat \sigma^y \Leftrightarrow \ket{11}$ and $\hat \sigma^z \Leftrightarrow \ket{10}$, then applying $\hat \sigma_x$ corresponds to flipping the replica spin in the Bell pair, while $\hat \sigma_z$ to flipping the physical spin. 
Therefore, the transition rule for the VMC in the doubled Hilbert space consists of flipping either a replica spin or two distinct physical spins randomly chosen. 

\begin{figure}
    \centering \includegraphics[width=0.8\linewidth]{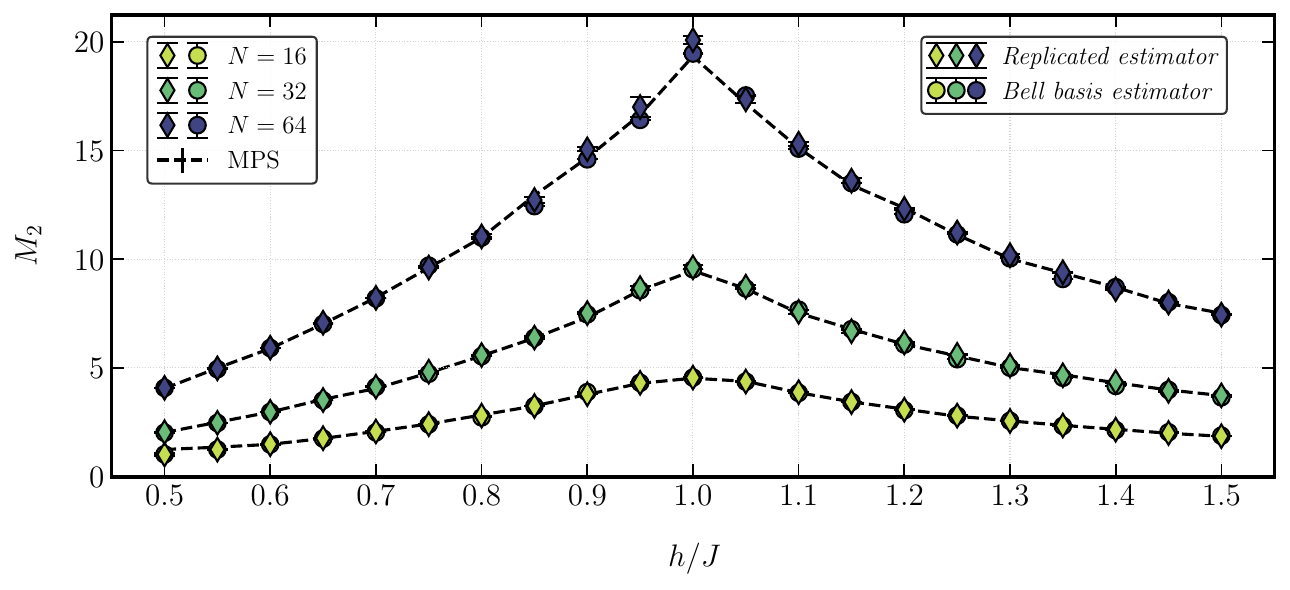}
    \caption{SRE $M_2$ in the ground state of the 1D TFI model for different values of $h/J$ and various system sizes $N$. 
    The circular markers correspond to values computed using the replicated estimator, while the diamond markers employing the Bell basis estimator. 
    For the Bell basis estimator $N_s = 2^{14} = 16 384$ samples are used, while for the replicated estimator $N_s = 2^{20} \approx 10^6$ for $N=16, 32$ and $N_s = 2^{30} \approx 10^9$ for $N=64$. 
    The dashed line is the reference calculation from perfect sampling with MPS~\cite{Lami2023} using $N_s =10^5$ samples and bond dimension $\chi=400$.}
    \label{fig:magic_ising}
\end{figure}

\cref{fig:magic_ising} demonstrates a strong agreement between the two Monte Carlo estimates and a reference calculation performed with MPS for the non-stabilizerness in the ground state of the 1D TFI model, even for the larger size attained. 
The statistical errors on the replicated estimator are generally larger than for the Bell basis estimator, due mostly to the presence in the sampling of outliers that are generated by the non-locality of the transformation $\hat{U}$. 
For this reason, a larger sample set is employed in the case of the replicated estimator. 

\subsection{Optimization of the Deep ViT wave function \label{sec:deep_vit}}
The Deep ViT ansatz $\ket{\Psi_{\theta}}$ used for approximating the ground state of the 1D and 2D $J_1$-$J_2$ Heisenberg model is the one introduced in~\cite{viteritti2023transformer_2d}.
The only difference lies in the feed-forward neural network of the encoder block, whose hidden dimension is doubled and the activation function is taken to be the GELU. 
The architecture employed in the simulations of this paper has the following structure: the hidden dimension is $d=60$, the number of heads is $h=10$, the linear size of the patches is $b=4$ in 1D and $b=2$ in 2D, and the number of layers is $n_l=4$ in 1D and for the $6 \times 6$ lattice while $n_l=8$ for the $8 \times 8$.
The attention weights are taken to be translational invariant, in order to encode the translational symmetry between the patches.
On top of this, the translational invariance over the whole lattice is imposed through quantum number projection inside the patches, by considering the symmetrized state:
\begin{equation}
\label{eq:translational}
    \Tilde{\Psi}_{\theta}(\sigma) = \sum_{\boldsymbol{r}} e^{-i \boldsymbol{k} \cdot \boldsymbol{r}}\Psi_{\theta}(T_{\boldsymbol{r}} \sigma), 
\end{equation}
where $T_{\boldsymbol{r}}$ is the translation operator of lattice vector $\boldsymbol{r}$ and $\boldsymbol{k}$ is a momentum vector. 
The sum in~\cref{eq:translational} runs over any possible lattice vector inside a patch, so it contains $b$ terms in 1D and $b^2$ terms in 2D.
The following table reports the number of variational parameters in the ViT architectures used in our simulations. 

\begin{table}[ht]
\centering
\begin{tabular}{|c|c|c|c|c|}
\hline
 & $N=32$ & $N=64$ & $N=6 \times 6$ & $N=8 \times 8$ \\
\hline
\# parameters & 154\,940 & 155\,260 & 154\,980 & 302\,540 \\
\hline
\end{tabular}
\caption{Parameter count for the ViT models employed in the simulations of the 1D and 2D $J_1$-$J_2$ Heisenberg model.
}
\label{tab:nqs_vs_mps}
\end{table}

The deep learning formulation of Stochastic Reconfiguration~\cite{rende2024simple,chen2024empowering} generates the following update for the parameters $\theta$: 
\begin{equation}
\label{eq:minsr}
    \delta \theta = \tau X (X^T X + \lambda I_{2N_s})^{-1} f.
\end{equation}

The matrix $X \in \mathbb{R}^{N_p \times 2 N_s}$, where $N_p$ is the number of parameters and $N_s$ the number of samples employed, is formed by the concatenation of the real and the imaginary part of the centered rescaled Jacobian $Y_{\alpha i} = (O_{\alpha i} - \Bar{O}_{\alpha}) / \sqrt{N_s}$ where $O_{\alpha i}  = \partial_{\theta_\alpha} \log \Psi_{\theta}(\sigma_i)$ and $\Bar{O}_{\alpha} = \sum_{i=1}^{Ns} O_{\alpha i} / N_s$.
Similarly, the vector $f \in \mathbb{R}^{2 N_s}$ is the concatenation between the real and minus the imaginary part of the centered rescaled local energy $\varepsilon_i = -2 [E_{\text{loc}}(\sigma_i) - \Bar{E}_{\text{loc}}]^* / \sqrt{N_s}$ where $E_{\text{loc}}(\sigma_i) = \bra{\sigma_i} \hat{H} \ket{\Psi_{\theta}} / \bra{\sigma_i} \ket{\Psi_{\theta}}$ and $\Bar{E}_{\text{loc}} = \sum_{i=1}^{Ns} E_{\text{loc}}(\sigma_i) / N_s$.
In our simulations, we adopt a cosine schedule for the learning rate $\tau$ from 0.03 to  0.005, $\lambda = 10^{-4}$ and $N_s = 6 \times 10^3$, in line with~\cite{rende2024simple}.
We first perform the VMC optimization without the translational symmetry until convergence, then we impose the symmetry and we continue the training for the same number of steps.
Since the $J_1$-$J_2$ Heisenberg Hamiltonian satisfies the SU(2) spin symmetry, the VMC sampling can be limited to the sector of total zero $z$-magnetization, thereby the transition rule consists of exchanges between nearest- or next-nearest neighbor spins.
We remark that the optimizations are realized without the Marshall sign rule prior~\cite{marshall1955antiferromagnetism}.

\subsection{Variational error on the $m_2$ \label{sec:variational_error}}
Let the converged NQS ansatz after the VMC optimization be written as:
\begin{equation}
\ket{\Psi_{\theta}} = \cos{\alpha}\ket{E_0} + \sin{\alpha}\ket{E_{>0}},
\end{equation}
with $\ket{E_0}$ and $\ket{E_{>0}}$ being respectively the ground state and a combination of the low-lying excited states of $\hat H$ (for simplicity we can consider only the first excited state) and $\alpha \in \mathbb{R}$. 
We have $\braket{E_0}{E_{>0}}=0$ as well as $\braket{E_0}{E_0}=\braket{E_{>0}}{E_{>0}}=1$. We note that, since $\sin^2\alpha + \cos^2 \alpha = 1$, also $\ket{\Psi_{\theta}}$ has unit norm. 

The energy variance on $\ket{\Psi_{\theta}}$ can be written as:
\begin{equation}
\label{eq:variance}
\begin{split}
\text{Var}(\hat{H}) &= \bra{\Psi_{\theta}}\hat{H}^2\ket{\Psi_{\theta}} -\bra{\Psi_{\theta}}\hat{H}\ket{\Psi_{\theta}}^2 = \\
    &= E^2_0\cos^2{\alpha} + E^2_{>0}\sin^2{\alpha} -E^2_0\cos^4{\alpha} -E^2_{>0}\sin^4{\alpha} -
    - 2 E_0E_{>0}\cos^2{\alpha}\sin^2{\alpha} =\\
    % &\approx \frac{(E_0-E_A)^2}{N^2} \left[\alpha^2 -\frac{4}{3}\alpha^4 \right] = \frac{(E_0-E_A)^2}{N^2} \alpha^2 + O(\alpha^4),
    &= (E_0-E_{>0})^2 \alpha^2 + O(\alpha^4), 
\end{split}
\end{equation}
where $E_0 = \bra{E_0} \hat{H} \ket{E_0}$, $E_{>0} = \bra{E_{>0}} \hat{H} \ket{E_{>0}}$ and the approximation is valid for $\alpha \approx 0$ (meaning that we have approximately converged to $\ket{E_0}$, as expected). 
For a gapped physical system, we can conclude from~\cref{eq:variance} that $\alpha \sim \sqrt{\text{Var}(\hat H)}$.

In the same fashion, the magic density $m_2$ on $\ket{\Psi_{\theta}}$ can be expressed to the lowest order in $\alpha$ as:
\begin{equation}
\label{eq:magic_density}
\begin{split}
    m_2(\ket{\Psi_{\theta}}) &= - \frac{1}{N}\log \bigg(\frac{1}{2^N}\sum_{ \hat{P} \in \mathcal{P}_N} \bra{\Psi_{\theta}}\hat{P} \ket{\Psi_{\theta}}^4 \bigg) = \\
    &= - \frac{1}{N}\log \bigg(\frac{1}{2^N}\sum_{\hat{P} \in \mathcal{P}_N}\left[ \cos^2{\alpha} \bra{E_0}\hat{P} \ket{E_0}+\sin^2{\alpha} \bra{E_{>0}}\hat{P} \ket{E_{>0}} + 2 \sin{\alpha}\cos{\alpha} \Re(\bra{E_0}\hat{P} \ket{E_{>0}})\right]^4\bigg) = \\
    &= -\frac{1}{N}\log \bigg({\frac{1}{2^N}\sum_{\hat{P}  \in \mathcal{P}_N}\bra{E_0}\hat{P} \ket{E_0}^4}\bigg)-\frac{8}{N}\bigg[\frac{\sum_{\hat{P}  \in \mathcal{P}_N}\bra{E_0}\hat{P} \ket{E_0}^3\Re(\bra{E_0}\hat{P} \ket{E_{>0}})}{\sum_{\hat{P}  \in \mathcal{P}_N}\bra{E_0}\hat{P} \ket{E_0}^4}\bigg]\alpha + O(\alpha^2) = \\
    % &+\bigg[2 \frac{\sum_P\left[\bra{E_{>0}}P\ket{E_0}^2+2\bra{E_{>0}}P\ket{E_0}\bra{E_0}P\ket{E_{>0}}+\bra{E_0}P\ket{E_{>0}}^2-2\bra{E_{>0}}P\ket{E_{>0}}\bra{E_0}P\ket{E_0}+2\bra{E_0}P\ket{E_0}^2\right]}{N \sum_P\bra{E_0}P\ket{E_0}^2}\bigg]\alpha^2 =
    &= m_2(\ket{E_0}) - \frac{8}{N} \bigg[\frac{\sum_{\hat{P}  \in \mathcal{P}_N}\bra{E_0}\hat{P} \ket{E_0}^3\Re(\bra{E_0}\hat{P} \ket{E_{>0}})}{\sum_{\hat{P} \in \mathcal{P}_N}\bra{E_0}\hat{P} \ket{E_0}^4}\bigg]\alpha + O(\alpha^2). 
\end{split} 
\end{equation}
  
This gives a way to estimate the error committed on $m_2$ due to the variational approximation of the ground state $\ket{E_0}$, namely $\Delta m_2 = |m_2(\ket{\Psi_{\theta}}) - m_2(\ket{E_0})|$. 
Indeed, considering that the term in the square brackets in~\cref{eq:magic_density} is $O(1)$ in $N$, we obtain:
\begin{equation}
\Delta m_2 \sim \frac{\sqrt{\text{Var}(\hat{H})}}{N}.
\end{equation}

\end{widetext}
\end{document}